\documentclass{aa}  

\usepackage{graphicx}
\usepackage{txfonts}
\usepackage{lipsum}
\usepackage{subcaption}
\usepackage{lscape}
\usepackage{placeins}
\usepackage[colorlinks,allcolors=blue]{hyperref}
\usepackage[normalem]{ulem}
\usepackage{orcidlink}
\usepackage{booktabs}
\usepackage{stfloats}

\newcommand\arcdeg{\mbox{$^\circ$}}%
\begin{document}
   \title{Ionized Gas Kinematics and Star-forming Clumps in Extremely Metal-poor Galaxies}

   \titlerunning{Kinematics and metallicity of XMP galaxies}
\author{
   Jyoti Yadav\orcidlink{0000-0002-5641-8102}\inst{1,2}
   \corrauth{yadavjyoti636@gmail.com}
    \and
    Casiana Muñoz Tuñón'\orcidlink{0000-0001-8876-4563}\inst{1, 2}
              \and  Cecilia Farina \orcidlink{0000-0003-4940-3751} \inst{1,2,3}
              \and  Alejandro Lumbreras-Calle \orcidlink{0000-0002-6696-7834}\inst{4}
              \and Jorge Sánchez Almeida \orcidlink{0000-0003-1123-6003}\inst{1, 2}
              \and Bruce G. Elmegreen \orcidlink{0000-0002-1723-6330} \inst{5}
}

\institute{
    Instituto de Astrofísica de Canarias, Vía Láctea s/n, E-38205 La Laguna, Spain
            \and
            Departamento de Astrofísica, Universidad de La Laguna, E-38206 La Laguna, Spain   
            \and 
            Isaac Newton Group of Telescopes, Apartado 321, E-38700 Santa Cruz de La Palma, Tenerife, Spain
            \and 
            Centro de Estudios de Física del Cosmos de Aragón (CEFCA), Unidad Asociada al CSIC, Plaza San Juan 1, 44001 Teruel, Spain 
            \and 
            Katonah, NY 10536, USA
}
\date{Received September 30, 20XX}
  \abstract
   {}
   {Extremely Metal Poor (XMP) galaxies are the least chemically evolved objects in the local Universe and provide important constraints on galaxy formation in low-metallicity environments. This work aims to investigate the ionised gas kinematics and star formation properties of XMP galaxies to identify signatures of disturbed or multi-component gas motions and to explore their connection with ongoing star formation, thereby probing the assembly and growth of low-mass galaxies.  
   } 
   {We analyse long-slit spectroscopic observations of nine XMP galaxies obtained with the Intermediate Dispersion Spectrograph on the 2.5\,m Isaac Newton Telescope. The study is primarily based on the H$\alpha$ emission line, which is used as a tracer of the ionised gas component. The kinematic measurements are used to identify ordered rotation, disturbed motions, and localized deviations associated with star-forming regions. Emission-line ratios are used to derive gas-phase metallicities, thereby enabling analysis of the chemical properties of the ionised gas.   
   }
   {The kinematic analysis reveals that a subset of galaxies exhibit ordered rotation, with the kinematic centres not always coincident with the luminous star-forming heads. Most systems show disturbed, asymmetric, or declining velocity profiles, while localized distortions near the bright star-forming heads suggest that massive star-forming clumps can influence the surrounding gas dynamics. We derive dynamical masses for the dominant clumps and investigate the mass--metallicity and size--metallicity relations. The clumps exhibit low metallicities consistent with the XMP regime and broadly comparable to those observed in low-mass dwarf galaxies. Three galaxies show counter-rotating or kinematically decoupled ionised-gas components, indicating the presence of multiple kinematic structures. These features, together with the low metallicities and intense star formation in the bright head regions, point to complex gas accretion and redistribution, while internal processes such as outflows, mixing, and stochastic star formation may also contribute to the observed properties. 
   }
   {The XMP galaxies in our sample are predominantly isolated systems, with metal-poor star formation concentrated in compact head regions. Three galaxies exhibit counter-rotating or kinematically distinct ionised-gas components, indicating complex gas kinematics and possible external gas acquisition. Such gas may be supplied through merger or interaction-driven processes, or through the accretion of relatively metal-poor gas from the surrounding intergalactic medium and cosmic filaments. These processes may contribute to the evolution of XMP galaxies, although internal mechanisms such as gas mixing, outflows, recycling, and stochastic star formation may also shape their observed properties.
   }
   \keywords{Galaxies: dwarf, Galaxies: evolution, Galaxies: kinematics and dynamics, Galaxies: abundances
    }
\maketitle
\nolinenumbers

\section{Introduction}

Galaxy evolution is closely tied to the progressive chemical enrichment of the interstellar medium (ISM), driven by metals produced through stellar nucleosynthesis and dispersed by supernova feedback. The gas-phase metallicity of a galaxy is regulated by the interplay between internal enrichment, metal loss via galactic outflows, and the accretion of lower-metallicity gas from the intergalactic medium \citep{Tinsley1980FCPh....5..287T, Edmunds1990MNRAS.246..678E,
Peroux2020ARA&A..58..363P}. This balance gives rise to well-established correlations between metallicity and global galaxy properties, such as stellar mass and star formation rate, as encapsulated in the mass--metallicity and fundamental metallicity relations \citep{Andrews2013ApJ...765..140A, Curti2020MNRAS.491..944C, SanchezAlmeida2014A&ARv..22...71S}. These scaling relations provide key constraints on the chemical evolution of galaxies across cosmic time \citep{Maiolino2019A&ARv..27....3M}.

Extremely metal-poor (XMP) galaxies are commonly defined as systems with gas-phase metallicities below one-tenth of the solar value ($Z<0.1,Z_{\odot}$; \citealt{Kunth2000A&ARv..10....1K}). Adopting a solar oxygen abundance of $12+\log(\mathrm{O/H})=8.69$ \citep{Asplund2009ARA&A..47..481A, Asplund2021A&A...653A.141A}, this corresponds to $12+\log(\mathrm{O/H})<7.69$. XMP galaxies are predominantly low-mass systems, with typical stellar masses of M$_{\star}$ $\sim$ 5 $\times$ 10$^{7}$ M$_{\odot}$ and elevated specific star-formation rates, with a median $\mathrm{sSFR}\sim10^{-8} \mathrm{yr}^{-1}$ \citep{Zou2024ApJ...961..173Z, Fukushima2025ApJ...988...46F}. Importantly, their low metallicities cannot be attributed solely to their low stellar masses. While dwarf galaxies generally follow the mass--metallicity relation, XMP galaxies exhibit metallicities significantly below those expected for their stellar masses, marking them as chemically anomalous systems. Their extreme metal deficiency may therefore reflect additional processes affecting their chemical evolution, such as the accretion of metal-poor gas, gas dilution, or interactions and mergers.

Despite their significance, XMP galaxies remain among the rarest star-forming systems in the local Universe. Early studies reported only a small number of confirmed XMP galaxies \citep{Kunth2000A&ARv..10....1K}. A systematic search based on the Sloan Digital Sky Survey (SDSS) Data Release 7 identified only $\sim$ 130 XMP galaxies, corresponding to nearly 0.1\% of nearby galaxies or $\sim$ 0.2\% considering only emission-line galaxies \citep{MoralesLuis2011ApJ...743...77M}. The subsequent SDSS-based studies have further expanded the census by identifying additional candidates and refining selection criteria \citep{SanchezAlmeida2016ApJ...819..110S}. Owing to their extremely low chemical enrichment, these systems are considered the closest local analogues of chemically primitive low-mass galaxies in the early Universe. However, identifying their counterparts at higher redshift remains challenging due to observational limitations. Consequently, systematic efforts have focused on increasing the sample size, characterizing the most metal-deficient systems, and extending such searches across all redshifts.
More recent efforts, combining wide-field photometric surveys such as Subaru/Hyper Suprime-Cam with spectroscopic data from the Dark Energy Spectroscopic Instrument (DESI) and SDSS DR17, have increased the sample to several hundred confirmed XMP galaxies and candidates \citep{Kojima2020ApJ...898..142K, Zou2024ApJ...961..173Z, Cheng2025MNRAS.540..128C}.

A significant fraction of XMP galaxies ($\sim$ 75\%) display irregular optical morphologies, often characterised by cometary/tadpole or multi-knot, asymmetric morphology \citep{Papaderos2008A&A...491..113P, MoralesLuis2011ApJ...743...77M}. In contrast, such morphologies are rare among the general population of star-forming galaxies; for example, only $\sim$ 0.2\% of UV-bright systems in the Kiso survey exhibit cometary shapes \citep{Miyauchi2010PNAOJ..13....9M, Elmegreen2012ApJ...750...95E}. These tadpole galaxies are characterised by a prominent, massive star-forming clump (M$_{\star \ clump}$ $\sim$ 10$^{7}$\, M$_{\odot}$) located at one end, followed by an elongated, diffuse tail \citep{Elmegreen2012ApJ...750...95E}. In several nearby systems, these dominant clumps are off-centre from the kinetic centre and exhibit lower metallicities compared to the surrounding galaxy, indicating the likely accretion of metal-poor gas \citep{SanchezAlmeida2013ApJ...767...74S, SanchezAlmeida2014A&ARv..22...71S, Almeida2015ApJ...810L..15S, SanchezAlmeida2016ApJ...819..110S, OlmoGarcia2017ApJ...834..181O, Almeida2018ApJ...869...40S}. These compact clumps likely comprise a mixture of young stellar populations and gas in different phases, including ionised, neutral, and molecular gas. Comparable localized metallicity depressions have also been reported in massive (M$_{\star}$ $\sim$ 10$^{10}$\, M$_{\odot}$) star-forming galaxies at high redshift \citep{Cresci2010Natur.467..811C, Queyrel2012A&A...539A..93Q}.

Several physical mechanisms have been proposed to explain these asymmetric features. One possibility is that they represent edge-on, low-mass disks in an early evolutionary stage, where prominent star-forming regions are observed along the line of sight, potentially driven by gravitational instabilities within the disk \citep{Elmegreen2009ApJ...701..306E, Elmegreen2010ApJ...722.1895E}. Alternatively, asymmetries may arise from external perturbations, such as interactions or minor mergers with low-mass companions \citep{Straughn2006ApJ...639..724S}. Another scenario involves the self-propagation of star formation across a gas-rich system or within a chemically unevolved gas cloud \citep{Papaderos1998A&A...338...43P, Papaderos2008A&A...491..113P}. The infall of metal-poor gas from the surrounding environment has also been suggested as a trigger for localized starburst activity, leading to both morphological asymmetries and reduced metallicities \citep{SanchezAlmeida2013ApJ...767...74S, SanchezAlmeida2014A&ARv..22...71S, Almeida2015ApJ...810L..15S, SanchezAlmeida2016ApJ...819..110S}. Evidence for ongoing gas accretion is also supported by studies of blue compact dwarf (BCD) galaxies, where irregular and, in some cases, kinematically disturbed or non-rotating HI structures are observed  \citep{Brinks1988MNRAS.231P..63B, Taylor1996ApJS..107..143T, Thuan1997ApJ...489..623T, Zee1998AJ....116.1186V, Pustilnik2001AJ....121.1413P, Hoffman2003AJ....126.2774H}.

Observations of HI emission reveal that local BCDs and XMP galaxies are typically embedded in extended reservoirs and streams of neutral gas, often exceeding their optical sizes by factors of $\gtrsim 3$ \citep{Begum2006MNRAS.365.1220B, Young2003ApJ...592..111Y, Thuan2004AJ....128..617T, Chengalur2006MNRAS.371.1849C, Ekta2008MNRAS.391..881E,  Ekta2009MNRAS.397..963E, Walter2008AJ....136.2563W, Ekta2010MNRAS.403..295E, Lelli2012A&A...544A.145L, Lelli2014MNRAS.445.1694L, Lelli2014A&A...566A..71L, Beccari2014A&A...570A..78B, Filho2013A&A...558A..18F}. These HI structures frequently exhibit asymmetric morphologies and disturbed kinematics, particularly in their outer regions. 
Furthermore, XMP galaxies are preferentially located in low-density environments, with $\sim$ 60\% residing in underdense regions and $\sim$ 75\% associated with large-scale structures such as cosmic voids and sheets \citep{Filho2015ApJ...802...82F}. This supports scenarios involving ongoing cold gas accretion of low-metallicity gas, which may simultaneously fuel star formation and dilute the metal content \citep{Kewley2006AJ....131.2004K}. Since gas mixing and chemical homogenisation in galactic disks proceed on timescales comparable to a single rotation period \citep{deAvillez2002ApJ...581.1047D, Yang2012ApJ...758...48Y}, the observed chemical inhomogeneities imply a recent origin. This suggests that the gas has been accreted in the recent past, likely in the form of residual streams associated with ongoing cosmological accretion \citep{SanchezAlmeida2013ApJ...767...74S, SanchezAlmeida2014A&ARv..22...71S}. Observational evidence for gas accretion flows has been reported at high redshift, most notably through the detection of inverted metallicity gradients, where central star-forming regions exhibit lower metallicity than the outer parts of galaxies \citep{Cresci2010Natur.467..811C}. Additional support comes from studies of high-column density HI absorbers \citep[e.g.,][]{vandeVoort2012MNRAS.421.2809V}, which are consistent with ongoing gas inflow from the intergalactic medium. In cosmological simulations, metal-poor gas clumps accreted from the intergalactic medium trigger localized star formation within galaxies \citep{Ceverino2016MNRAS.457.2605C}.

In the local Universe, however, such accretion signatures are largely indirect and are primarily inferred from the properties of dwarf galaxies. In particular, several studies have shown that, at fixed stellar mass, galaxies with higher star formation rates tend to exhibit lower metallicities \citep{LaraLopez2010A&A...519A..31L, Mannucci2010MNRAS.408.2115M, Perez-Montero2013A&A...549A..25P}. Furthermore, quiescent BCD galaxies are found to be more metal-rich than their actively star-forming counterparts, suggesting that episodes of enhanced star formation are linked to the dilution of the interstellar medium by metal-poor gas \citep{SanchezAlmeida2008ApJ...685..194S, SanchezAlmeida2009ApJ...698.1497S}. Similar evidence has emerged from other compact, metal-poor star-forming systems, including extreme emission-line galaxies and blueberry galaxies, which exhibit low metallicities, high specific star formation rates, and strong nebular emission, consistent with recent episodes of metal-poor gas accretion fueling intense star formation \citep[e.g.,][]{Amorin2015yCat..35780105A, Yang2017ApJ...847...38Y, Lumbreras-Calle2022A&A...668A..60L}.

Overall, a detailed investigation of the dynamical properties and the distribution of stellar, ionised, and neutral gas components is essential to disentangle the origin of XMP galaxies morphologies and the physical processes governing their evolution. In particular, resolved gas kinematics can provide more direct evidence of external gas accretion through signatures such as disturbed velocity fields, kinematic decoupling, and metallicity variations associated with star-forming regions. Thus, in this work, we combine ionised gas kinematics and metallicity measurements of XMP galaxies to probe signatures of external gas accretion directly. Using spatially resolved H$\alpha$ velocity together with chemical abundances of the star-forming regions, we investigate kinematic disturbances and metallicity variations that may trace ongoing inflow. This enables us to assess how gas accretion influences star formation and the dynamical structure of these systems.


\begin{figure*}
    \centering
    \includegraphics[width=1\linewidth]{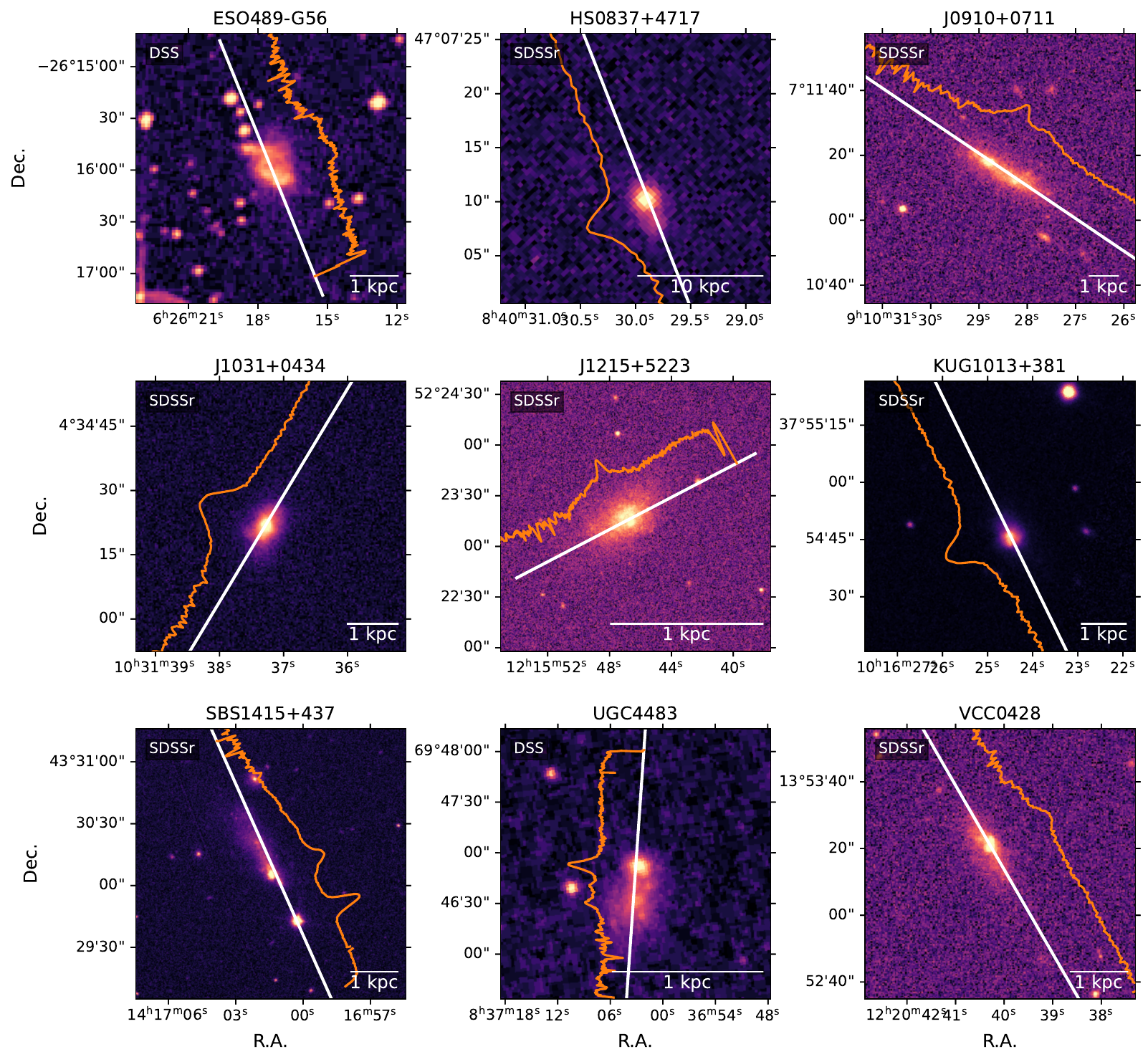}
    \caption{The images show the XMP galaxies from the SDSS (where available; \citealt{York2000AJ....120.1579Y}) and the Digitized Sky Survey (DSS; \citealt{STSci2020ipac.data.I441S}). The white line indicates the slit orientation used for the spectroscopic observations, while the orange curve represents the logarithmic intensity profile extracted from the INT long-slit data along the major axis of the galaxy.}
    \label{fig:slit_on_image}
\end{figure*}

\begin{table*}
\caption{Details of the XMP galaxies in this work.}                 
\label{table:detail_galaxy}    
\centering  
\resizebox{\textwidth}{!}{
\begin{tabular}{c c c c c c c c c c c}       
\hline\hline               
Galaxy Name & R.A & Dec. & D & scale & sin i & 12+log(O/H) & log M$_{*}$ & log M$_{HI}$   & log SFR &  Morphology \\
 & hh:mm:ss & dd:mm:ss & Mpc & pc/$\arcsec$ & &  & M$_{\odot}$ & M$_{\odot}$   & M$_{\odot}$ yr$^{-1}$ &   \\
 (1)& (2) & (3) & (4) & (5) & (6) & (7) & (8) & (9)  & (10) & (11)   \\
\hline                      
   ESO\,489-G56 & 06:26:17.50  & -26:15:56.80  & 4.23   & 20.5  & 0.80  & 7.49 & -- & 6.95  & -- &  -- \\
HS0837+4717  & 08:40:29.91  & +47:07:10.28  & 174.30 & 845.8 & 0.69  & 7.64 & 7.97 &  --  & 0.55 & cometary\\
J0910+0711   & 09:10:28.76  & +07:11:17.33  & 22.00  & 106.7 & 0.94  & 7.63 & -- & --  & -- & cometary\\
J1031+0434   & 10:31:37.28  & +04:34:21.53  & 18.10  & 87.7  & 0.70  & 7.70 & 6.61 & -- & -1.53 & cometary\\
J1215+5223   & 12:15:46.63  & +52:23:13.84  & 3.33   & 16.1  & 0.85  & 7.43 & 6.01 & 7.09 & -2.82 & multi-knot\\
KUG1013+381  & 10:16:24.50  & +37:54:45.85  & 19.90  & 96.5  & --    & 7.58 & 6.58 & 8.15 & -1.46 & cometary\\
SBS1415+437  & 14:17:01.44  & +43:30:06.13  & 11.66  & 56.5  & 0.57  & 7.43 & -- & 8.24 & -- & multi-knot\\
UGC\,4483    & 08:37:02.97  & +69:46:32.80  & 5.00   & 24.2  & 0.92  & 7.58 & -- & 7.88 & -- & --\\
VCC\,0428    & 12:20:40.28  & +13:53:22.42  & 13.10  & 63.5  & 0.91  & 7.64 & 6.20 & 7.42 & -1.92 & cometary\\
\hline                                  
\end{tabular}}
\tablefoot{Column 1: Source name. Column 2: Right Ascension. Column 3: Declination. Column 4: Hubble flow distance corrected for Virgocentric infall, from NED. Column 5: Physical scale corresponding to 1\arcsec at the distance of the galaxy. Column 6: Inclination angle of the source from \citet{Filho2013A&A...558A..18F}. Column 7: Metallicity, 12 + log (O/H), from \citet{MoralesLuis2011ApJ...743...77M}. Column 8: Logarithm
of the photometric stellar mass from SDSS, Max-Planck Institute for Astrophysics, Johns Hopkins
University (MPA–JHU) catalog \citep{Kauffmann2003MNRAS.341...33K, Brinchmann2004MNRAS.351.1151B}. Column 9: Logarithm of the HI mass taken from \citet{Filho2013A&A...558A..18F}. Column 10: Logarithm of the star-formation rate from the SDSS MPA-JHU catalog \citep{Kauffmann2003MNRAS.341...33K, Brinchmann2004MNRAS.351.1151B}. Column 11: SDSS optical morphology adapted from \citet{MoralesLuis2011ApJ...743...77M} and \citet{Filho2013A&A...558A..18F}.}
\end{table*}

\begin{figure}
    \centering
    \includegraphics[width=1\linewidth]{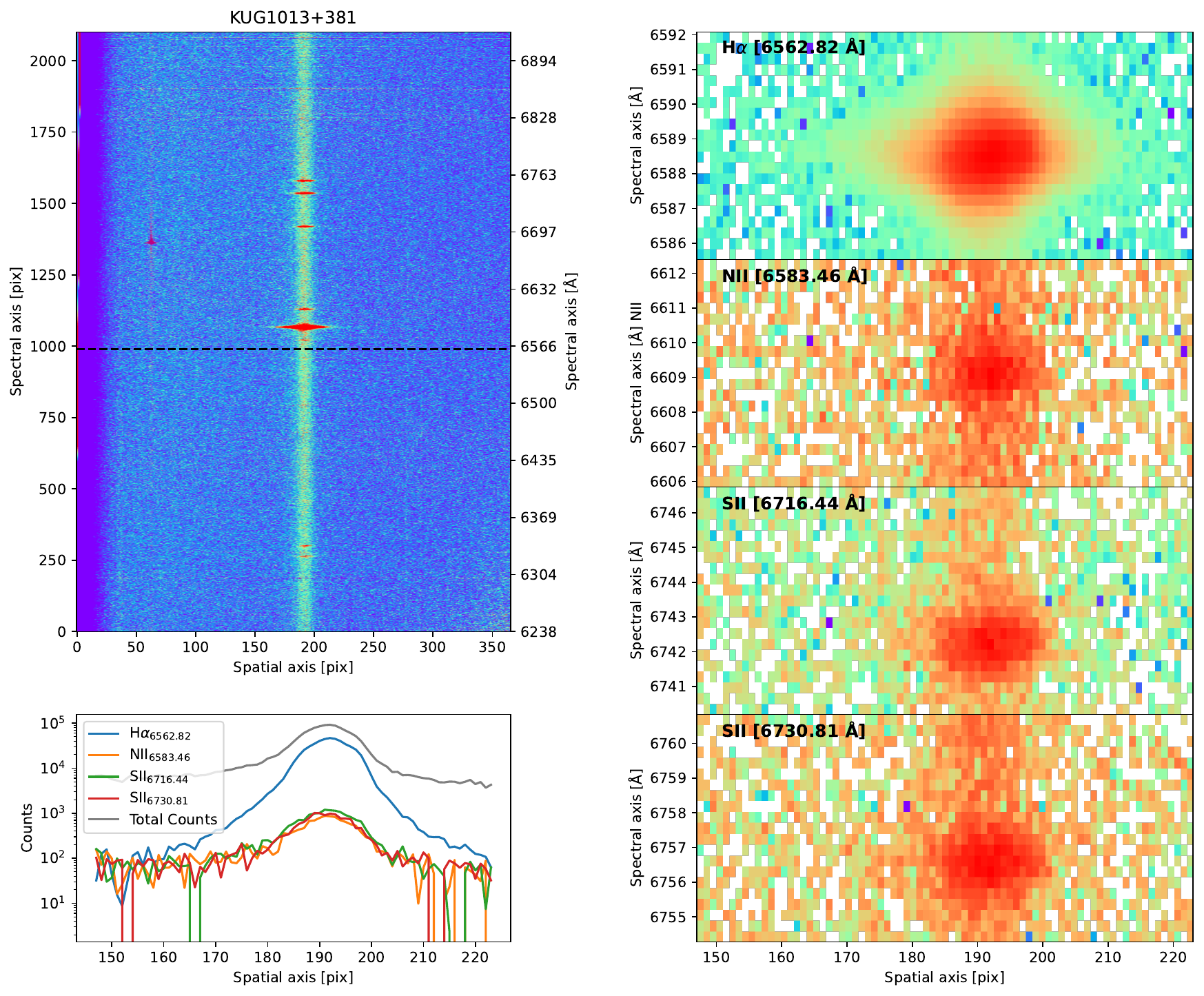}
    \caption{The top-left panel shows the slit image from the observations of the XMP galaxy KUG1013+381 as an example. The lower-left panels present the intensity profiles obtained by integrating the spectrum over wavelength, together with the emission-line profiles of H$\alpha$, [\ion{N}{ii}], [\ion{S}{ii}]$\lambda6716$, and [\ion{S}{ii}]$\lambda6730$. The right panel shows a zoomed-in view of the spatial distribution of the corresponding emission lines along the major axis of the galaxy.}
    \label{fig:65B}
\end{figure}

\begin{figure*}
    \centering
    \includegraphics[width=1\linewidth]{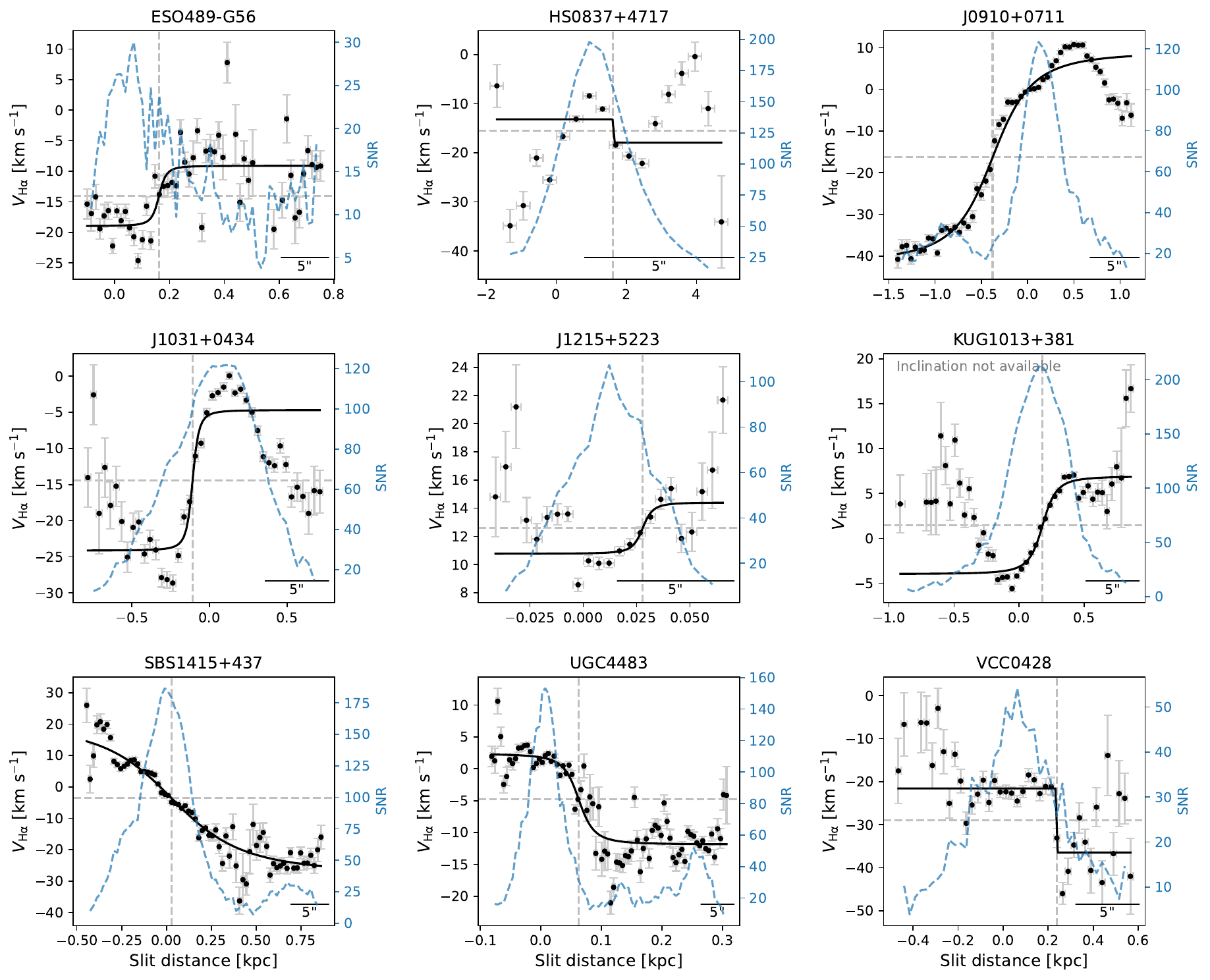}
    \caption{Position--velocity curves of the nine XMP galaxies. The horizontal axis represents the projected distance along the major axis of the galaxy, measured from the position of the tadpole head, which is the brightest H$\alpha$ region in each system. The horizontal scale bar in each panel corresponds to 5\arcsec. The filled circles show the data points with errors, and the solid black curve shows the best-fitting analytic rotation curve to the observed velocity measurements. The grey dashed lines indicate the systemic velocity (horizontal) and the kinematic centre derived from the fit (vertical). The dashed light blue curve shows the S/N of H$\alpha$ emission flux.}
    \label{fig:rotation_curve}
\end{figure*}

\begin{figure*}
    \centering
    \includegraphics[width=1\linewidth]{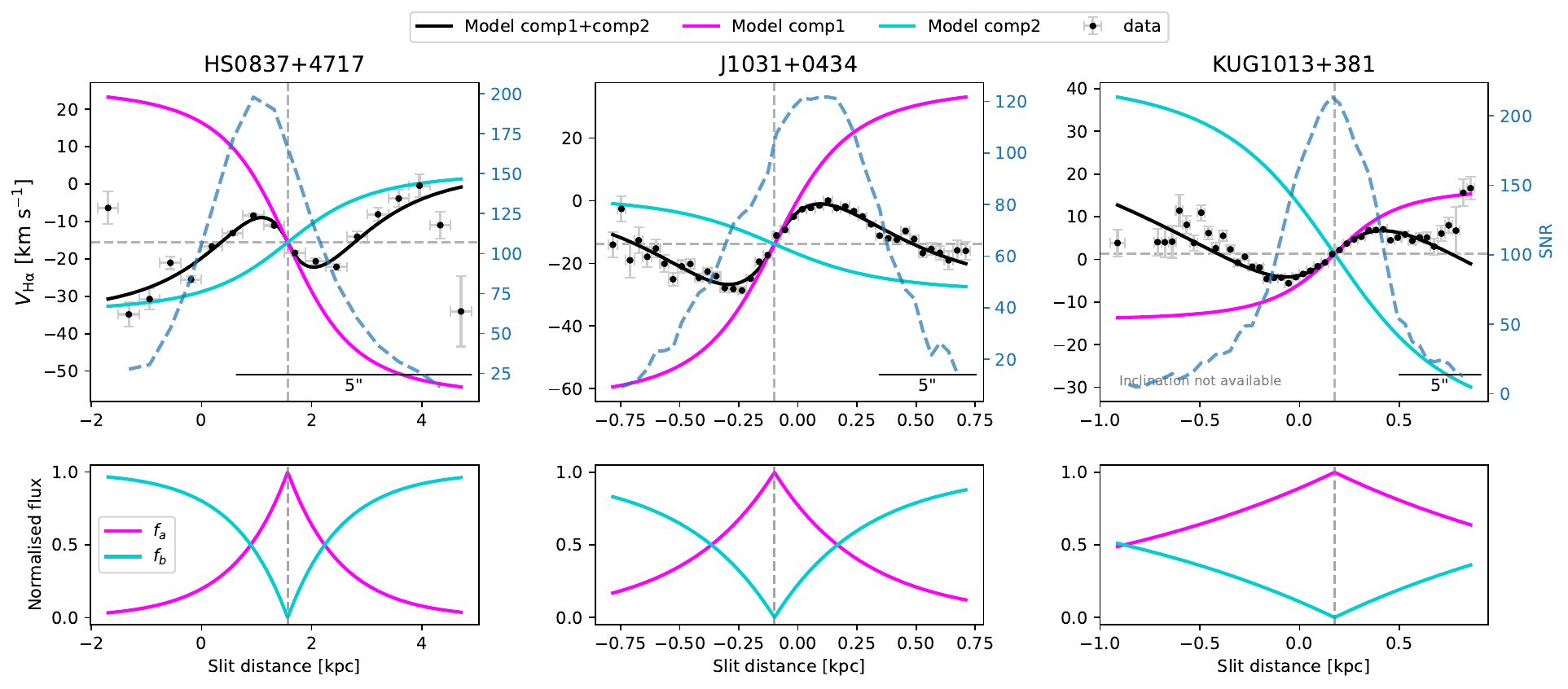}
    \caption{Top panels show the position--velocity profiles of the galaxies fitted with a two-component rotation-curve model. The magenta and cyan curves represent the two kinematic components, while the black curve shows their combined model. Bottom panels
    show the flux variation of the component with distance along the slit in magenta and cyan, respectively. The overall presentation follows the same scheme as Figure~\ref{fig:rotation_curve}.}
    \label{fig:two_component}    
\end{figure*}

\begin{figure*}
    \centering
    \includegraphics[width=1\linewidth]{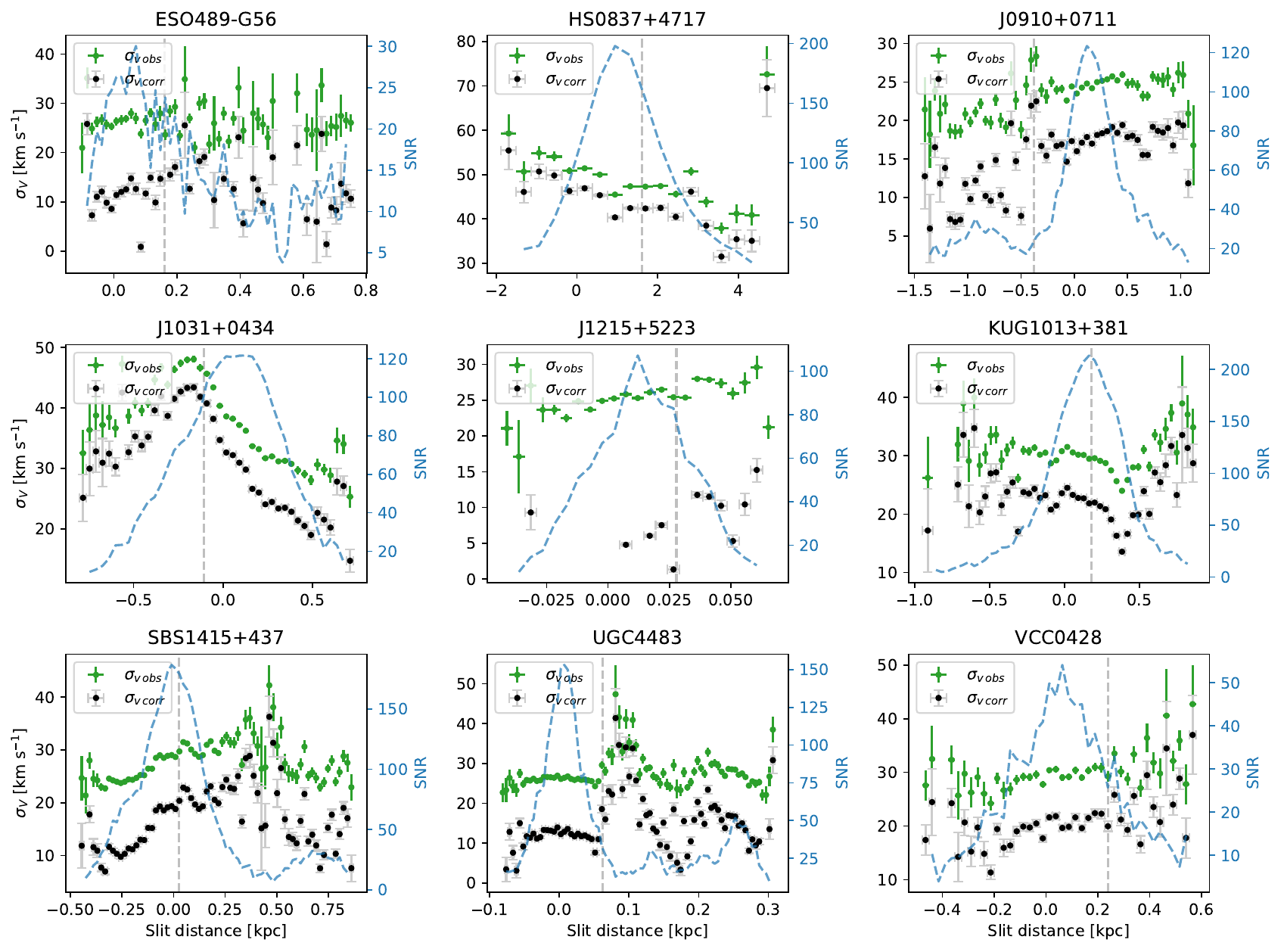}
    \caption{Velocity dispersion ($\sigma_{V}$) profiles of the galaxies along the slit. The blue curve and grey vertical line show the S/N of H$\alpha$ and the kinematic centre of the galaxy, respectively. The green and black points represent the observed and corrected velocity dispersions, respectively.}
    \label{fig:vel_disp}
\end{figure*}

\begin{figure}
 \centering
    \includegraphics[width=1\linewidth]{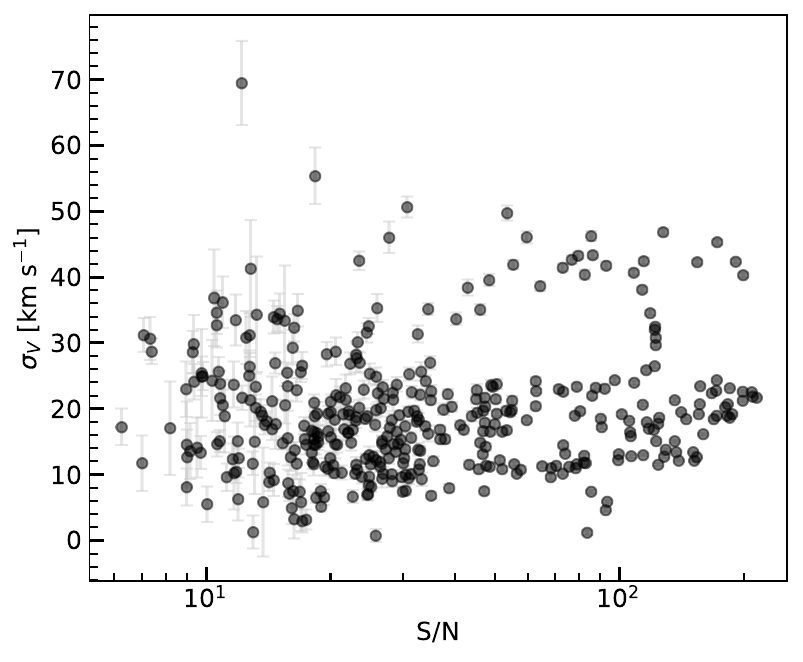}
    \caption{Velocity dispersion as a function of the H$\alpha$ emission-line S/N for all spatial bins in the sample galaxies.}
    \label{fig:SNR}
\end{figure}

\begin{figure*}
    \centering
    \includegraphics[width=1\linewidth]{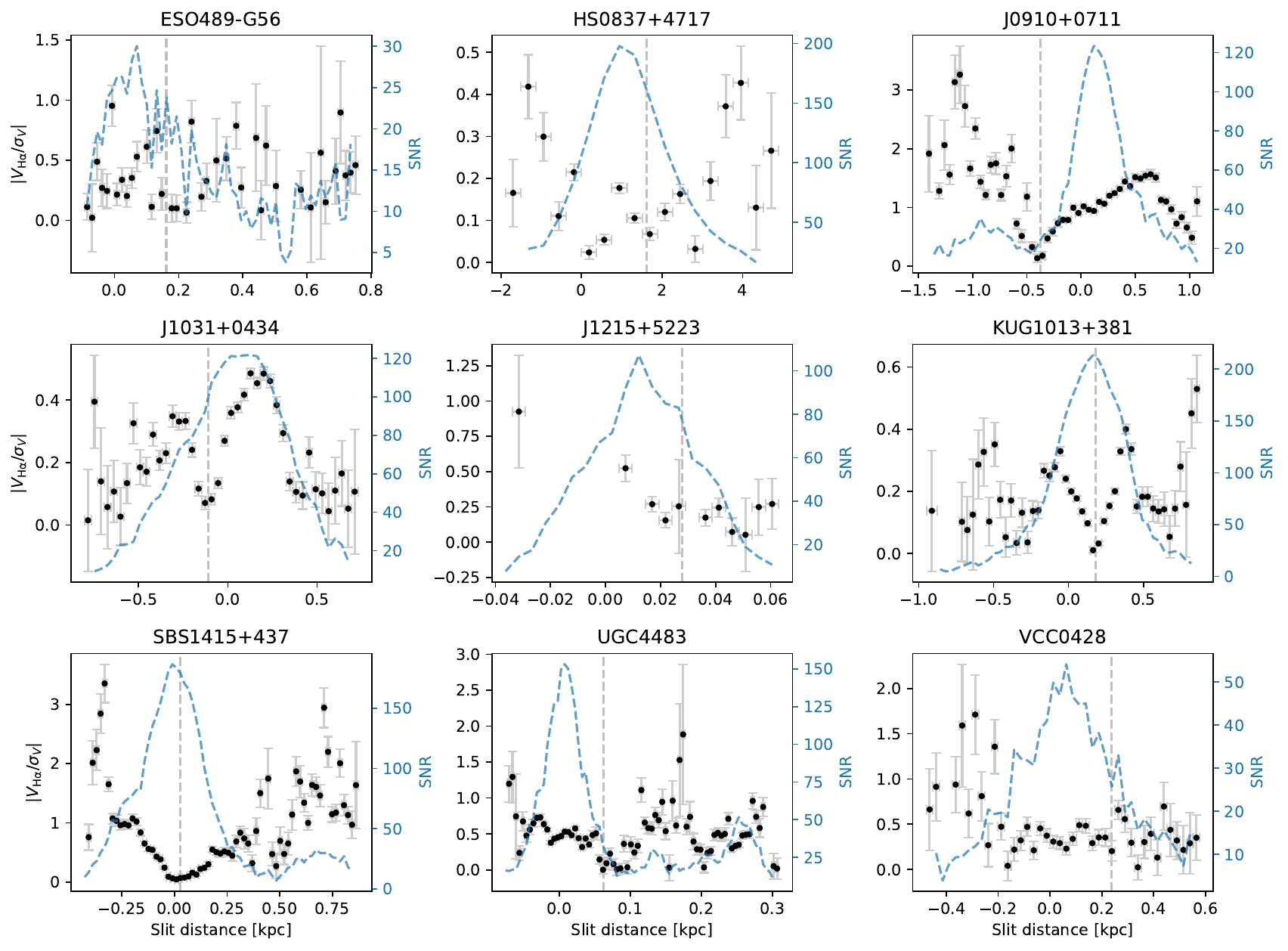}
    \caption{V$_{\textrm{H}\alpha}$/$\sigma_{V}$ profiles of the galaxies (black). The blue curve and grey vertical line show the S/N of H$\alpha$ and the kinematic centre of the galaxy.}
    \label{fig:vel_by_sigma}
\end{figure*}

\begin{figure*}
    \centering
    \includegraphics[width=1\linewidth]{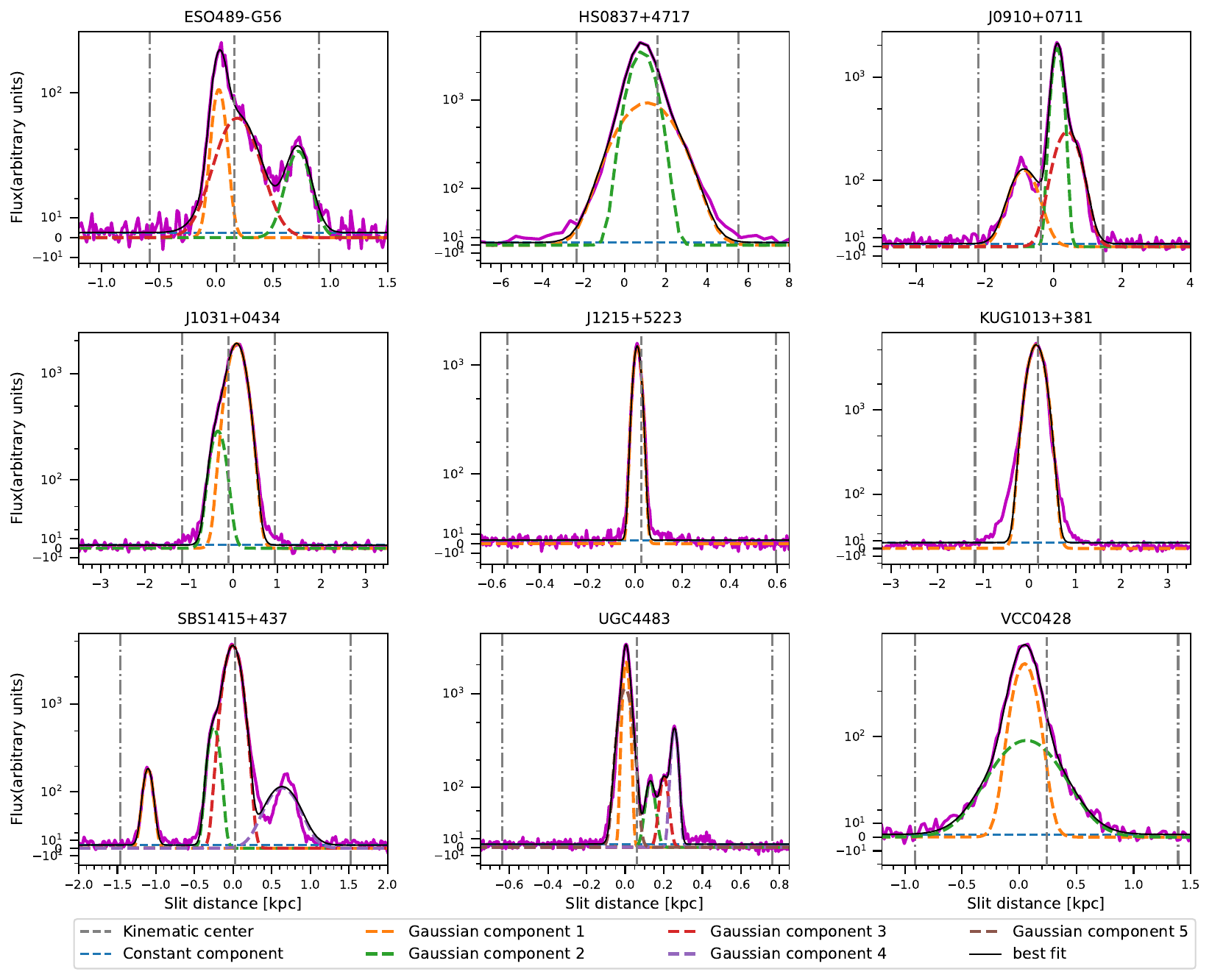}
    \caption{H$\alpha$ intensity profiles of the galaxies measured along the slit (magenta). The individual Gaussian components used in the fit are shown separately, while the combined best-fitting model, obtained from the sum of all Gaussian components, is shown in black. The vertical dotted line marks the kinematic centre, and the dot-dashed lines indicate the spatial extent of the galaxy. The number of Gaussian components required to fit the H$\alpha$ emission-line profiles varies from one to five among the galaxies in the sample. The largest number of components (five) is fitted in UGC\,4483. The y-axis has asinh scaling ($y\rightarrow 50\, \text{asinh}^{-1}(y/50)$) for better visualisation.}
    \label{fig:intensity}
\end{figure*}


\section{Observations and data reduction}
We observe a sample of nine XMP galaxies to investigate their ionised gas kinematics and chemical properties. The nine XMP galaxies analysed in this work are selected from the sample of XMP galaxies compiled by \citet{MoralesLuis2011ApJ...743...77M}. The observations were obtained on 16 and 17 February 2016 using the Intermediate Dispersion Spectrograph (IDS) mounted on the 2.5\,m Isaac Newton Telescope (INT) at the Observatorio del Roque de los Muchachos \citep{Laing1985VA.....28..483L}. We used the IDS-Red+2 configuration with the H1800V grating centred on the H$\alpha$ emission line, which also includes the [\ion{N}{ii}]$\lambda6583$ transition. The observations employed the H1800V grating (1800\,lines\,mm$^{-1}$), providing a spectral dispersion of 0.34\,\AA\,pixel$^{-1}$. The detector has a plate scale of 0.44\arcsec\,pixel$^{-1}$. The instrumental setup covers a spectral range of approximately 700\,\AA\ centred on the selected wavelength, ensuring simultaneous coverage of H$\alpha$ and the [\ion{N}{ii}]$\lambda6583$ emission line used for the metallicity analysis. The spectral resolution, set by the 1\arcsec\ slit width, corresponds to 1.4 pixels or 0.49\,\AA. 

The alignment of the slit along the major axis of each galaxy is shown in Fig.~\ref{fig:slit_on_image}. The observational strategy was designed to probe the dynamical properties of the galaxies through long-slit spectroscopy. The slit was aligned along the prominent head--tail morphology of each source, allowing us to trace velocity gradients and construct rotation curves along the major axis. 
We observed five individual exposures for each galaxy, with an integration time of 1200s. The data reduction followed standard procedures, including bias subtraction, flat-field correction, cosmic-ray removal, and subtraction of sky emission lines. The observations were obtained under seeing conditions of approximately 1\arcsec. The individual exposures were aligned to a common reference frame and subsequently combined to produce a final image with an exposure time of 6000s (1200 $\times$ 5) and improved S/N. Figure~\ref{fig:65B} shows the combined image of one of the galaxies, KUG1013+381. The combined image corresponds to an exposure time of 6000s. Table~\ref{table:detail_galaxy} shows the details of the galaxies. The instrumental spectral resolution was determined from Gaussian fits to arc-lamp emission lines used for wavelength calibration. The average full width at half maximum (FWHM) of the arc-lamp emission lines was measured for each galaxy, with values ranging from 31 to 51 km\,s$^{-1}$.

\section{Analysis and results}
\subsection{Rotation Curves and dynamical mass estimation}
We derived the velocity curves along the major axes of the galaxies using the ionised gas kinematics traced by the H$\alpha$ emission line. We spatially binned the spectra along the slit direction to achieve an S/N of 3 in the H$\alpha$ line for reliable measurements in each bin, while pixels with S/N $>$ 3 remained unbinned. For each spatial bin, we modelled the underlying continuum emission using a first-order Chebyshev polynomial and subsequently subtracted the continuum. We modelled the H$\alpha$ emission line in each spatial bin using a single Gaussian profile to estimate the line centroid, FWHM, and amplitude. The centroid of the fitted profile was used to derive the radial velocity, while the FWHM was used to estimate the velocity dispersion of the ionised gas. We estimated the uncertainties on the fitted parameters using a Markov Chain Monte Carlo (MCMC) approach, providing statistically robust error estimates.

To characterise the rotation of the galaxies, we modelled the observed velocity profiles by fitting an analytic rotation curve. We employed a parametrisation motivated by the universal rotation curve (URC) formalism presented by \citet{Salucci2007MNRAS.378...41S}, which has been shown to successfully describe galaxy rotation curves over a broad range of masses. The model parameters were determined by fitting the function to the observed velocity data.

\begin{equation}
    V(d)= V_{0}+ V_{1} \frac{d-d_{0}}{\sqrt{\Delta^{2}+(d-d_{0})^2}}
\end{equation}

where $V(d)$ denotes the observed velocity at a projected distance $d$ along the slit. In this study, we use the H$\alpha$ emission line to trace the ionised-gas kinematics and denote the corresponding line-of-sight velocity as $V_{\textrm{H}\alpha}$. The parameters $V_{0}$, $V_{1}$, $\Delta$, and $d_{0}$ are treated as free parameters and are determined through the fitting procedure. In this formulation, $d_{0}$ defines the kinematic centre of the rotation curve, while $V_{1}$ represents the velocity amplitude, such that $V \rightarrow V_{0} \pm V_{1}$ as $d \rightarrow \pm \infty$. The parameter $\Delta$ represents the velocity gradient. We estimated the approximate dynamical masses of the galaxies from $V_{\textrm{H}\alpha}$, as described in Appendix~\ref{sec:dynamical_mass}. These estimates should be regarded as lower limits, as they are based on the observed H$\alpha$ extent and do not account for the contribution of asymmetric drift \citep{Read2016MNRAS.462.3628R, Oh2025Ap&SS.370..129O}.

Figure~\ref{fig:rotation_curve} presents the inclination-corrected rotation curves for the nine galaxies in our sample. The radial extent of the rotation curve (R$_{\textrm{H}\alpha}$) along the slit is restricted to data points with a bin size $<10$ pixels and uncertainties in $V_{\textrm{H}\alpha}$ below $20~\mathrm{km s^{-1}}$. The galaxy inclinations, derived from the optical morphologies, are adopted from \citet{Filho2013A&A...558A..18F}, who estimated the inclination angles assuming thick circular disks. We note, however, that XMP galaxies are often intrinsically triaxial rather than axisymmetric, with intermediate-to-major axis ratios of $B/A \approx 0.7$ and minor-to-major axis ratios of $C/A \approx 0.15$--0.4, the latter increasing toward lower-mass systems \citep{Putko2019ApJ...883...10P}. Consequently, the inclination-corrected rotation velocities should be regarded as approximate, as deviations from the assumed geometry may introduce systematic uncertainties in the inferred inclinations.

The dashed blue curve shows the H$\alpha$ signal-to-noise ratio (S/N), estimated as the ratio of the emission-line flux to the standard deviation of the continuum. The S/N profile traces the relative distribution of H$\alpha$ emission along the slit, providing a qualitative indication of the locations of active star-forming regions with respect to the kinematic centre and the underlying gas kinematics. The reference position (zero point) along the spatial axis was defined using the brightest region of each galaxy, which coincides with the dominant star-forming clump. The rotation curves do not reach a well-defined outer plateau, suggesting that the full extent of the kinematic disk is not traced by the observations. This is likely because H$\alpha$ emission primarily originates from young, massive star-forming regions and may not extend to the outermost regions of the galaxy. The vertical dashed lines mark the kinematic centres, inferred from the fitted rotation models. The kinematic centres do not coincide with the photometric peaks, indicating a spatial offset between the dominant star-forming region and the kinematic centres of galaxies.

Three galaxies (J0910+0711, SBS1415+437, and UGC 4483) exhibit velocity gradients consistent with ordered rotation, although their velocity fields also show minor but measurable perturbations of approximately $\sim$ 5 km s$^{-1}$. The remaining systems display disturbed or asymmetric velocity fields. The minor perturbations in the velocity field are spatially coincident with the bright, compact star-forming heads, indicating a direct link between intense star formation and local gas kinematics. These deviations appear as localized departures from the large-scale velocity gradient, suggesting a dynamical origin. The concentration of young, massive stars in these regions implies that stellar feedback, including supernova explosions, stellar winds, and radiation pressure, may inject energy and momentum into the surrounding interstellar medium, producing localized perturbations in the velocity field on sub-kiloparsec scales. In addition, these star-forming heads may correspond to massive, self-gravitating clumps with significant internal dynamics, potentially including independent rotation. Similar massive, gravitationally bound star-forming clumps have been identified in the XMP galaxy KISO\,5639 \citep{Elmegreen2016ApJ...825..145E, Elmegreen2018ApJ...859L..22E}. Owing to their substantial masses, these clumps can contribute significantly to the local gravitational potential, modifying the enclosed mass distribution and producing localized changes in the gas velocity field near the clump boundaries. In such cases, the star-forming regions constitute dynamically important substructures rather than passive tracers of the host gravitational potential, and their kinematics may be misaligned with or distinct from the global rotation of the galaxy \citep[e.g.,][]{Immeli2004ApJ...611...20I, Moiseev2012MNRAS.423.1831M, Wisnioski2012MNRAS.422.3339W, Moiseev2015MNRAS.449.3568M, Cava2018NatAs...2...76C}.

\subsection{Two-Component Rotation Curve Fitting }
The rotation curves of the sample galaxies exhibit localized velocity reversals and kinematic irregularities that cannot be reproduced by a single rotating-disk model. The mismatch between the observations and the best-fitting single-component model indicates that these features are intrinsic to the ionised gas kinematics rather than arising from noise or limited spatial resolution. The observed velocity fields, therefore, require a more complex description involving multiple kinematic components. To account for this complexity, we model the rotation curves using a flux-weighted two-component model, in which each component is described by Equation~\ref{eq:constrined_same_center_same_velocity}. The two components are associated with the main clump, which rotates in one direction, and the outer regions, which rotate in the opposite direction. The flux weighting allows the relative contribution of each component to vary along the galaxy. We explored a range of model configurations, including both co-rotating and counter-rotating components, with the kinematic centres either treated as free parameters or constrained to coincide. The constrained model, in which both components share a common kinematic centre, provides the most satisfactory representation of the observed velocity profiles and successfully reproduces the observed kinematic features.

The two-component model provides a significantly improved representation of the observed velocity profiles in three galaxies (Fig.~\ref{fig:two_component}), successfully reproducing both the large-scale velocity gradients and the localized velocity reversals. Three galaxies, HS0837+4717, J1031+0434, and KUG1013+381, show signatures of ionised-gas components with opposite velocity gradients, indicating the presence of kinematically distinct gas with opposing angular momentum. Such signatures may arise from the accretion or inflow of gas whose angular momentum is misaligned with, or opposite to, that of the pre-existing disk. Given the collisional nature of the gas, interactions between the components can result in shocks and dissipation, and the observed configuration may therefore represent an ongoing or recent phase of accretion before the gas becomes dynamically settled. Other possibilities include spatially separated gas structures projected along the line of sight, or kinematic disturbances associated with galaxy interactions. Localised feedback processes, such as supernova-driven ejecta or AGN-driven outflows, may also perturb the ionised gas and contribute to complex or apparently reversed velocity components. Higher-resolution observations combining stellar and multi-phase gas kinematics will therefore be crucial for distinguishing externally accreted gas from feedback- or interaction-driven disturbances and for establishing the origin and persistence of the observed counter-rotating components. Overall, these systems exhibit significant kinematic complexity, indicating non-equilibrium gas dynamics in a substantial fraction of the sample.

\subsection{Velocity dispersion profiles}
We derive the non-thermal ionised gas velocity dispersion ($\sigma_{V}$) from the width of the H$\alpha$ emission line by fitting a single Gaussian profile to each spatial bin along the major axis. The observed width ($\sigma_{\mathrm{o}}$) is corrected for instrumental resolution ($\sigma_{\mathrm{i}}$), natural broadening ($\sigma_{\mathrm{n}}$) and thermal broadening ($\sigma_{\mathrm{th}}$)

\begin{equation}
\sigma_{V}^{2} = \sigma_{\mathrm{o}}^{2} - \sigma_{\mathrm{i}}^{2} - \sigma_{\mathrm{n}}^{2} - \sigma_{\mathrm{th}}^{2}
\end{equation}

The instrumental broadening was corrected using the widths of the arc-lamp emission lines measured for each galaxy. A uniform thermal broadening component of FWHM ($\mathrm{FWHM}_{\mathrm{th}} )= 25$ km\,s$^{-1}$ is adopted, corresponding to ionised gas at a characteristic temperature of $\sim 1.4 \times 10^{4}$ K, typical of HII regions \citep{Osterbrock1989agna.book.....O, Osterbrock2006agna.book.....O}. The natural line width FWHM ($\mathrm{FWHM}_{\mathrm{n}}$) of H$\alpha$ is negligible in comparison, contributing only $\sim 7$ km\,s$^{-1}$ \citep{Rozas2006A&A...455..539R}. 

Figure~\ref{fig:vel_disp} presents the resulting velocity dispersion profiles as a function of projected distance along the major axis for all galaxies in the sample. The S/N of the H$\alpha$ emission is shown as the dashed blue curve in each panel. The uncertainties increase toward the outer regions where the emission becomes faint, and the S/N decreases. In some galaxies (J0910+0711, J1031+0434, KUG1013+381, SBS1415+437, and UGC4483), elevated velocity dispersions are observed outside the main clump regions, albeit with larger associated uncertainties. Although these measurements have lower S/N, the $\sigma_{V}$ values remain broadly consistent with those derived in higher S/N regions (Fig.~\ref{fig:SNR}) and show little dependence on the adopted spatial binning, indicating that the observed enhancements are not dominated by noise or binning effects. We can resolve the H$\alpha$ and [N\,\textsc{ii}] lines separately (Fig.~\ref{fig:65B}), and [N\,\textsc{ii}] does not contaminate $\sigma_{V}$. The increased dispersion outside the star-forming clumps may instead reflect enhanced turbulence or unresolved kinematic substructure in these regions.

The corrected velocity dispersions span $\sim$ 10--50 km s$^{-1}$ across the sample, with values of $\sim$ 10--20 km s$^{-1}$ in the star-forming regions, consistent with previous measurements of HII regions \citep{Terlevich1981MNRAS.195..839T, TenorioTagle1993ApJ...418..767T, MunozTunon1996AJ....112.1636M, Fuentes2000AJ....120..752F, Telles2001ApJ...548..671T}. Since the kinematics are traced using H$\alpha$ emission associated with HII regions, the observed velocity dispersions likely reflect a combination of gravitational and stellar feedback-driven processes. Possible contributors include virialized gas motions within star-forming complexes, turbulent stirring by young stellar populations, and the cumulative effect of unresolved expanding shells and filaments driven by stellar winds and supernova explosions.
\citep{Terlevich1981MNRAS.195..839T, TenorioTagle1993ApJ...418..767T, Dyson1979A&A....73..132D, TenorioTagle1996ApJ...456..264T, Melnick1999MNRAS.302..677M}.

In ESO489-G56, UGC4483, and VCC0428, the velocity dispersion remains approximately constant across the star-forming clumps, whereas in HS0837+4717, J0910+0711, J1031+0434, J1215+5223, KUG1013+381, and SBS1415+437 it exhibits a systematic gradient across the clumps. The enhanced velocity dispersion may arise from turbulent gas in the inter-clump medium, where multiple unresolved kinematic components along the line of sight broaden the observed emission-line profiles through the superposition of distinct velocity fields \citep{Melnick1999MNRAS.302..677M, MunozTunon1996AJ....112.1636M}. Similar trends were identified by \citet{OlmoGarcia2017ApJ...834..181O}, who classified the velocity-dispersion profiles of XMP star-forming regions into distinct categories and suggested that increasing velocity dispersion toward the galaxy center may arise from enhanced turbulence associated with the interaction of the starburst gas with the surrounding interstellar medium, possibly reflecting the inward migration of massive star-forming clumps driven by tidal forces. In contrast, nearly constant velocity dispersions across the clumps are consistent with coherent kinematic structures that follow the local gas motion.

Figure~\ref{fig:vel_by_sigma} shows the $V_{H\alpha}$/$\sigma_{V}$ profiles along the major axis, where $V_{H\alpha}$ is the measured rotation velocity and $\sigma_{V}$ is the local velocity dispersion derived from the H$\alpha$ emission line. Most galaxies exhibit $V_{H\alpha}$/$\sigma_{V}$ values below unity over a substantial fraction of their extent, indicating that random and turbulent gas motions are comparable to or dominate over ordered rotation. Such low $V_{H\alpha}$/$\sigma_{V}$ values suggest that the ionised gas is dynamically disturbed and not fully supported by regular disk rotation \citep{Binney1978MNRAS.183..501B}. Instead, the gas kinematics appear strongly influenced by turbulence, non-circular motions, and localized dynamical perturbations. Such kinematics are not uncommon among low-mass dwarf galaxies, many of which are observed to be dispersion-supported or only weakly rotationally supported \citep{Kirby2014MNRAS.439.1015K, Leaman2012ApJ...750...33L, Kirby2017ApJ...834....9K, Wheeler2017MNRAS.465.2420W}. This behaviour may reflect their formation as intrinsically thick, dynamically hot systems rather than rotationally supported disks \citep{Wheeler2017MNRAS.465.2420W}. Alternatively, tidal stirring and past gravitational interactions can reduce $V_{H\alpha}$/$\sigma_{V}$ by dynamically heating the stellar and gaseous components, although the efficiency of this process depends strongly on the orbital configuration and interaction history \citep{Mayer2001ApJ...547L.123M, Mayer2001ApJ...559..754M, Kazantzidis2011ApJ...726...98K, Lokas2015ApJ...810..100L}.

\begin{table*}
\caption{Physical parameters of Galaxies.}                 
\label{table:clump_properties}   
\centering 
\resizebox{\textwidth}{!}{
\begin{tabular}{lcccccccccc}
\hline
Galaxy & Clump & $R_{\rm clump}$ & log($M_{\rm clump}$) & $R_{\rm gal}$ & $R_{\rm H\alpha}$ & log($M_{dyn \thinspace approx}$) & log($M_{\rm dyn}$) & $R_c/R_g$ & $M_c/M_{\rm dyn}$ & 12+log(O/H) \\
& No. & kpc & M$_{\odot}$  & kpc & kpc & M$_{\odot}$  & M$_{\odot}$ &  &  & 
 \\ 
 (1) & (2) & (3) &(4) & (5)& (6) & (7)  & (8) & (9) & (10) & (11)
 \\ 
\hline
ESO489-G56 & 1 & 0.07 $\pm$ 0.0 & 7.1 $\pm$ 0.04 & 0.74 & 0.59 $\pm$ 0.01 & 6.52 $\pm$ 0.01 & 7.65 & 0.1 & 0.928 & -- \\
 & 2 & 0.13 $\pm$ 0.01 & 7.19 $\pm$ 0.12 &  &  &  &  & 0.172 & 0.939 & -- \\
 & 3 & 0.22 $\pm$ 0.01 & 7.58 $\pm$ 0.04 &  &  &  &  & 0.296 & 0.991 & -- \\
HS0837+4717 & 1 & 1.39 $\pm$ 0.02 & 9.25 $\pm$ 0.01 & 3.93 & 3.31 $\pm$ 0.0 & 6.64 $\pm$ 0.0 & -- & 0.354 & -- & 7.73 $\pm$ 0.03 \\
 & 2 & 0.48 $\pm$ 0.0 & 8.79 $\pm$ 0.01 &  &  &  &  & 0.122 & -- & 7.77 $\pm$ 0.03 \\
J0910+0711 & 1 & 0.47 $\pm$ 0.02 & 7.89 $\pm$ 0.05 & 1.82 & 1.5 $\pm$ 0.04 & 8.33 $\pm$ 0.01 & -- & 0.256 & -- & 8.0 $\pm$ 0.66 \\
 & 2 & 0.15 $\pm$ 0.0 & 7.63 $\pm$ 0.01 &  &  &  &  & 0.085 & -- & 7.72 $\pm$ 0.03 \\
 & 3 & 0.42 $\pm$ 0.01 & 8.08 $\pm$ 0.02 &  &  &  &  & 0.229 & -- & 7.72 $\pm$ 0.03 \\
J1031+0434 & 1 & 0.22 $\pm$ 0.0 & 8.2 $\pm$ 0.01 & 1.05 & 0.82 $\pm$ 0.02 & 7.26 $\pm$ 0.01 & -- & 0.214 & -- & 7.46 $\pm$ 0.03 \\
 & 2 & 0.19 $\pm$ 0.01 & 8.33 $\pm$ 0.02 &  &  & &  & 0.181 & -- & -- \\
J1215+5223 & 1 & 0.02 $\pm$ 0.0 & 6.12 $\pm$ 0.01 & 0.56 & 0.07 $\pm$ 0.0 & 4.72 $\pm$ 0.02 & 7.69 & 0.032 & 0.796 & 7.68 $\pm$ 0.07 \\
KUG1013+381 & 1 & 0.19 $\pm$ 0.0 & 7.88 $\pm$ 0.01 & 1.36 & 1.09 $\pm$ 0.01 & 6.87 $\pm$ 0.0 & 8.3 & 0.137 & 0.949 & 7.43 $\pm$ 0.03 \\
SBS1415+437 & 2 & 0.08 $\pm$ 0.0 & 7.07 $\pm$ 0.02 & 1.49 & 0.83 $\pm$ 0.02 & 7.95 $\pm$ 0.01 & 7.52 & 0.051 & 0.94 & 7.56 $\pm$ 0.1 \\
 & 3 & 0.11 $\pm$ 0.0 & 7.54 $\pm$ 0.01 &  &  &  &  & 0.072 & 1.003 & 7.49 $\pm$ 0.03 \\
 & 4 & 0.24 $\pm$ 0.03 & 7.75 $\pm$ 0.06 &  &  &  &  & 0.164 & 1.031 & -- \\
UGC4483 & 1 & 0.02 $\pm$ 0.0 & 6.47 $\pm$ 0.01 & 0.7 & 0.24 $\pm$ 0.0 & 6.45 $\pm$ 0.01 & -- & 0.023 & -- & 7.43 $\pm$ 0.03 \\
 & 2 & 0.03 $\pm$ 0.0 & 6.8 $\pm$ 0.07 &  &  &  &  & 0.039 & -- & -- \\
 & 3 & 0.03 $\pm$ 0.0 & 6.81 $\pm$ 0.09 &  &  &  &  & 0.036 & -- & -- \\
 & 4 & 0.02 $\pm$ 0.0 & 6.61 $\pm$ 0.02 &  &  &  &  & 0.027 & -- & -- \\
 & 5 & 0.04 $\pm$ 0.0 & 6.8 $\pm$ 0.01 &  &  &  &  & 0.05 & -- & 7.45 $\pm$ 0.04 \\
VCC0428 & 1 & 0.12 $\pm$ 0.0 & 7.63 $\pm$ 0.02 & 1.15 & 0.7 $\pm$ 210 & 6.96 $\pm$ 2.48 & 8.58 & 0.108 & 0.889 & -- \\
 & 2 & 0.35 $\pm$ 0.01 & 8.07 $\pm$ 0.03 &  &  &  &  & 0.303 & 0.941 & -- \\
\hline
\end{tabular}
}
\tablefoot{Column (1): Source name. Column (2): Number of detected clumps. Column (3): Clump radius (kpc). Column (4): Logarithm of the clump mass. Column (5): Galaxy radius adopted from NED, corresponding to the SDSS isophotal radius in r band. Column (6): Extent of the H$\alpha$ emission in kpc. Column (7): Logarithm of the galaxy dynamical mass estimated from the rotation velocity measured over the observed H$\alpha$ extent. Column (8): Logarithm of the dynamical mass from \citet{Filho2013A&A...558A..18F}. Column (9): Ratio of clump radius to the extent of galaxies R$_{gal}$. Column (10): Ratio of clump mass to dynamical mass from \citet{Filho2013A&A...558A..18F}. Column (11): Gas-phase metallicity of the clump. For some galaxies, the \ion{N}{II} emission is too weak (S/N $< 3$) to derive reliable metallicity estimates for all identified clumps. Note: Clump 1 in SBS1415+437 is identified as a foreground star and has therefore been excluded from the table.}
\end{table*}

\subsection{Clump mass estimation}
The dynamical masses of the star-forming clumps are estimated from the observed emission-line widths under the assumption of virial equilibrium. If the isotropic velocity distribution that balances the clump gravity has a dispersion $\sigma$, then the corresponding mass is estimated following the formulation of \citet{Bosch2009AJ....137.3437B, SanchezAlmeida2013ApJ...767...74S}. We note, however, that some clumps may not be in perfect virial equilibrium due to the effects of stellar feedback, turbulence, or ongoing gas accretion, and the derived dynamical masses should therefore be regarded as approximate.

\begin{equation}\label{eq:clump_mass}
M = (1.20 \times 10^{5}\,M_{\odot})\,R_{\mathrm{e}}\,\mathrm{FWHM}^{2},
\end{equation}
where the effective radius $R_{\mathrm{e}}$ is the half-light radius of the clump. The velocity dispersion is related to the FWHM through

\begin{equation}
\begin{split}
\mathrm{FWHM} &= 2\sqrt{2\ln 2}\,\sigma_{o-i-n} \\
\sigma_{o-i-n}&= \sigma_{\mathrm{o}}^{2} - \sigma_{\mathrm{i}}^{2} - \sigma_{\mathrm{n}}^{2} 
\end{split}
\end{equation}

The measured line widths ($\sigma_{o-i-n}$) are corrected for natural and instrumental broadening to recover the intrinsic velocity dispersion of the ionised gas. We do not correct for thermal broadening, as both thermal and non-thermal motions contribute to the pressure support against gravity and should therefore be included in the virial mass estimation of the clumps.

The effective radius $R_{\mathrm{e}}$ is derived by fitting a one-dimensional Gaussian profile to the H$\alpha$ intensity distribution of clumps along the major axis of the galaxy. The FWHM is twice the half-light radius. Figure~\ref{fig:intensity} presents the Gaussian fit applied to the emission profile of the clump. The measurements are corrected for atmospheric seeing effects using the equation below,

\begin{equation}
R_{\mathrm{e}}^{2} = R_{\mathrm{e,0}}^{2} - \left({S}/{2}\right)^{2},
\end{equation}
where $R_{\mathrm{e,0}}$ is the observed effective radius and $S$ is the seeing.

The virial masses estimated from the velocity dispersions represent the total enclosed dynamical mass within the clump radius. They therefore include the contribution of all gravitating components (gas, stars, and dark matter). The derived properties of the star-forming clumps are presented in Table~\ref{table:clump_properties}. These clumps span a substantial range in both size and mass. The estimated clump radii span approximately $\sim$ 20pc--1kpc, while their masses range from $\sim$ 10$^{6}$ to $10^{9}$ M$_{\odot}$. This broad distribution reflects the diversity of star-forming structures present in the sample, from compact, low-mass regions to more extended and massive complexes. The clump masses are estimated using the mean velocity dispersion measured across each star-forming clump. However, the velocity dispersion is not spatially uniform across multiple star-forming clumps, and several clumps do not exhibit the constant dispersion profiles expected for fully virialized systems. Consequently, the derived dynamical masses should be regarded as approximate rather than precise virial masses. The galaxies HS0837+4717, J1031+0434, KUG1013+381, and VCC0428 are sufficiently compact that their star-forming regions remain unresolved and appear as a single clump, similar to the single-knot galaxies reported by \citep{Hinojosa2016A&A...592A.122H, Lumbreras-Calle2022A&A...668A..60L}. These unresolved structures may contain contributions from multiple individual clumps that cannot be distinguished at the available spatial resolution. Consequently, the measured intensity profiles represent the combined emission of these regions, potentially leading to overestimates of the individual clump sizes and masses. In particular, the limited resolution may result in higher inferred clump masses by blending multiple star-forming regions into a single big clump.

\begin{figure*}
    \centering
    \includegraphics[width=1\linewidth]{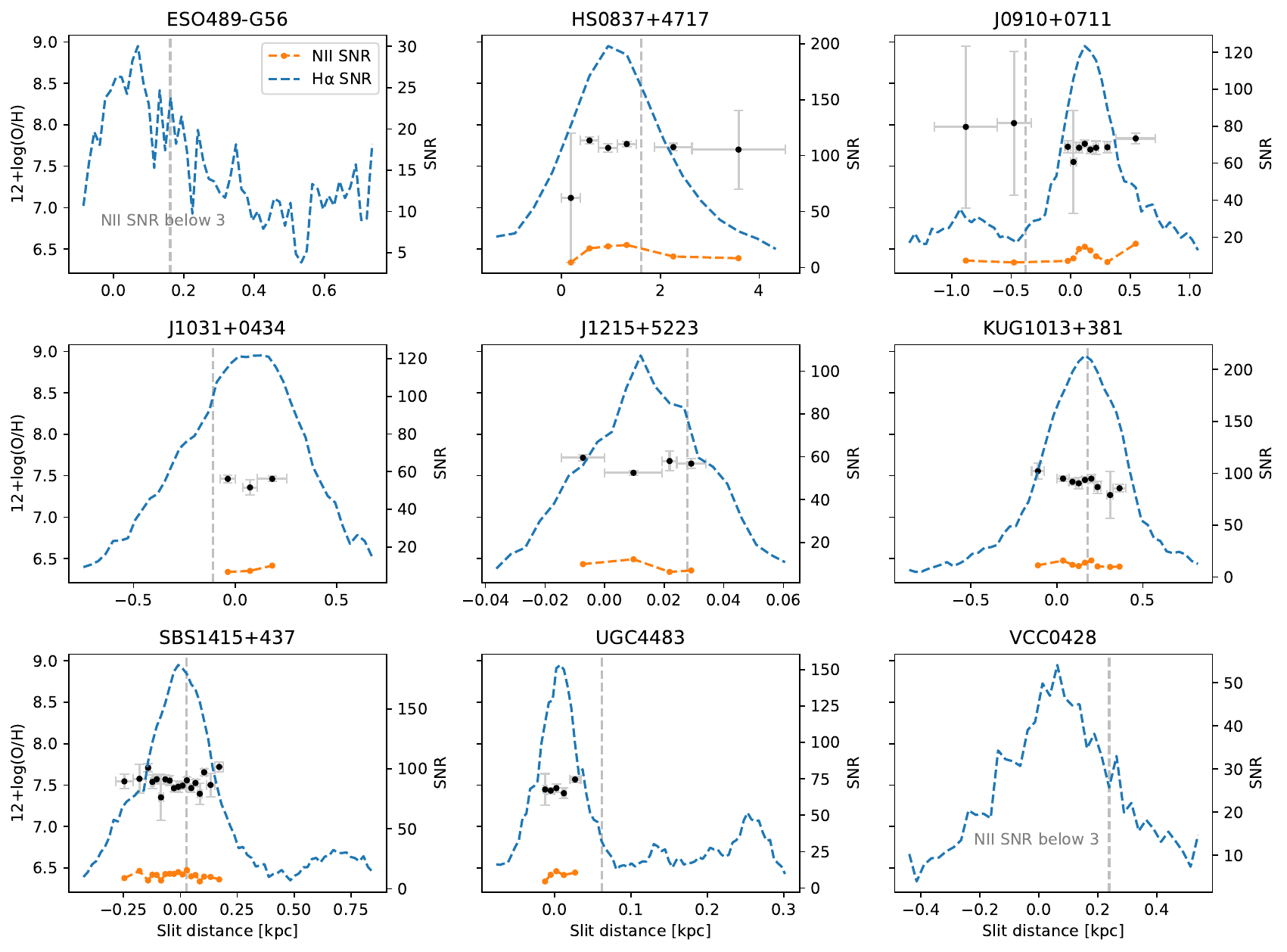}
    \caption{Metallicity profiles of the XMP galaxies derived from the [N\textsc{ii}] and H$\alpha$ emission-line ratio. The symbols show the measured values, with error bars representing the corresponding uncertainties. The light blue and orange curves show the S/N of H$\alpha$ and NII emissions, respectively. The vertical dotted line marks the kinematic centre.}
    \label{fig:metallicity_NII}
\end{figure*}

\begin{figure*}
    \centering
    \sidecaption
    \includegraphics[width=6cm]{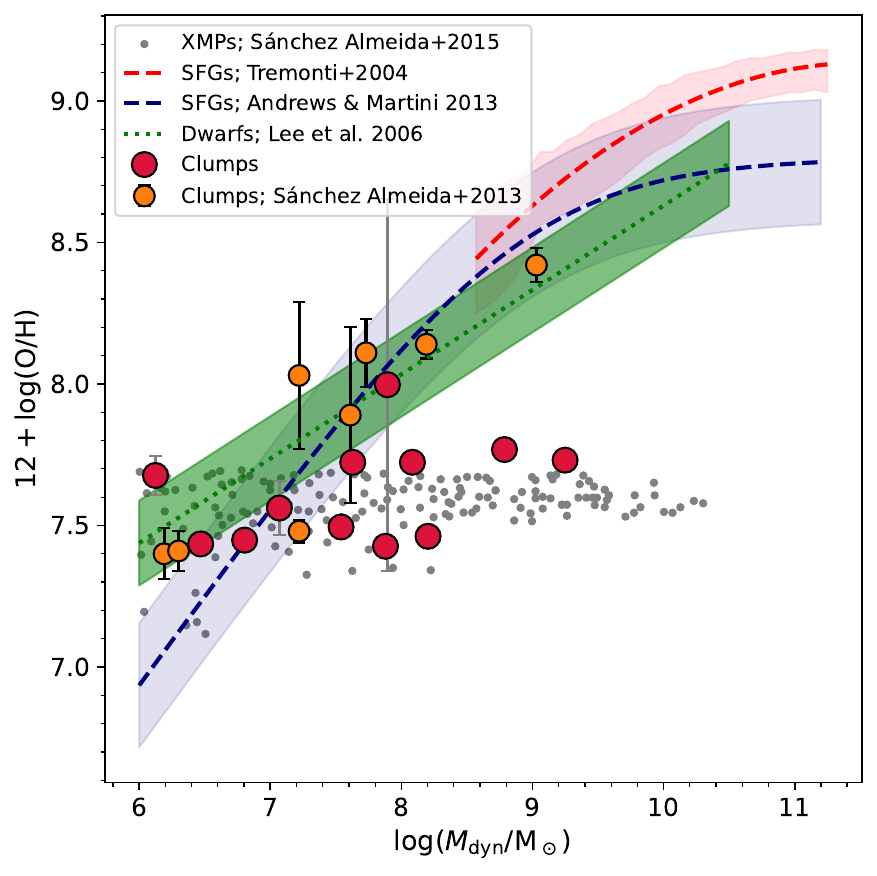} 
     \includegraphics[width=6cm]{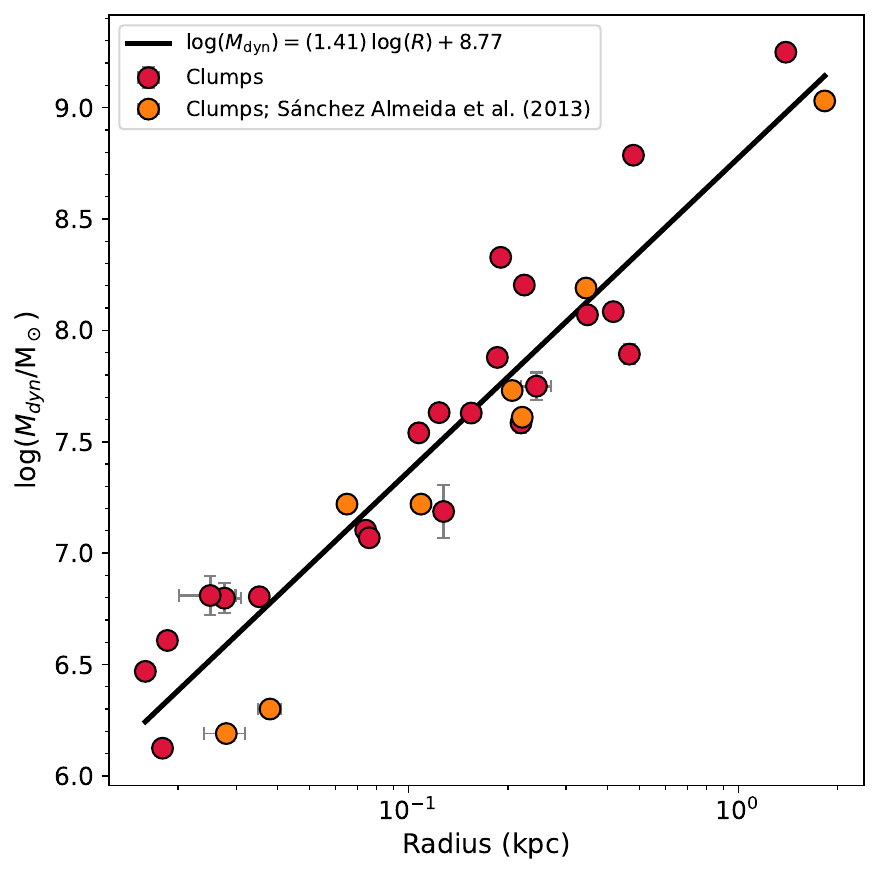} 
    \caption{Left: Dynamical mass–metallicity relation for the clumps (red) in this work. The orange points indicate clump properties from \citet{SanchezAlmeida2013ApJ...767...74S}. We also show the stellar mass–metallicity relations reported for low-mass galaxies (green line; \citet{Lee2006ApJ...647..970L}), high-mass star-forming galaxies (red line; \citet{Tremonti2004ApJ...613..898T}), and from \citet{Andrews2013ApJ...765..140A} in blue. Right: Dynamical mass–size relation for clumps (red) in this work, with orange points showing the clump measurements from \citet{SanchezAlmeida2013ApJ...767...74S}.}
    \label{fig:metallicity_clump}
\end{figure*}

\subsection{Metallicity}
Gas-phase metallicities are typically derived from emission-line diagnostics using either the direct (temperature-based) method \citep{Pagel1981ARA&A..19...77P, Osterbrock1989agna.book.....O} or empirical calibrations \citep{Shi2005A&A...437..849S, Kewley2008ApJ...681.1183K} based on strong emission-line ratios. Although the direct method is preferred whenever feasible, it requires measurements of emission-line fluxes spanning the optical to infrared regime, making it observationally expensive. For large samples or spectra with limited wavelength coverage, strong-line calibrations provide a practical alternative by linking easily accessible line ratios to oxygen abundance. In this work, we use a calibration based on the [\ion{N}{ii}]$\lambda6583$/H$\alpha$ flux ratio of \citet{PerezMontero2009MNRAS.398..949P}.

\begin{equation}
12 + \log(\mathrm{O/H}) = 9.07 + 0.79 \log \left( {[\mathrm{NII}]\,\lambda 6583}/{\mathrm{H}\alpha} \right)
\end{equation}

This method is particularly suitable for our data since both lines are simultaneously covered within the observed spectral range. This approach enables a consistent and reliable estimate of the gas-phase metallicity across the sample. We spatially binned the spectra along the slit direction by imposing a minimum S/N of 3 on the [\ion{N}{ii}] emission line, and applied the same binning to the H$\alpha$ emission. We subtracted the continuum emission and then fitted the emission lines in each spatial bin with single Gaussian profiles to derive their fluxes. 
Fig.~\ref{fig:metallicity_NII} shows the radial metallicity profiles derived along the major axis of each galaxy. Metallicity measurements are reliably obtained only in high-S/N star-forming regions, where the [N{\sc ii}] $\lambda6583$ emission line is detected with sufficient significance to enable robust line-ratio diagnostics. Figure~\ref{fig:metallicity_clump} presents the mass--metallicity relation (MZR) and the mass--size relation for the star-forming clumps identified in our sample. The clump metallicity is estimated by averaging the measurements over the clump diameter. For comparison, we show the MZR for massive star-forming galaxies from \citet{Tremonti2004ApJ...613..898T}, together with the low-mass galaxy relation from \citet{Lee2006ApJ...647..970L}. Since reliable metallicity measurements are not available at every position along the slit, the metallicity assigned to the clumps is calculated as the mean of the measurements at positions where the emission lines are sufficiently strong to provide reliable abundance estimates.

The clump metallicities are broadly consistent with the low-metallicity regime observed in low-mass dwarf galaxies and with those reported for star-forming regions in XMP dwarf galaxies by \citet{SanchezAlmeida2013ApJ...767...74S}. The mass--size relation was quantified by fitting a power-law model to the data. Assuming a relation of the form $M\propto R^{\alpha}$, we performed a least-squares linear regression in logarithmic space by fitting 

\begin{equation}
\log(M) = \alpha\,\log({R})+\beta
\end{equation}

where $M$ is the dynamical mass and $R$ is the effective radius. The fitted slope, $\alpha$, corresponds to the exponent of the mass--size scaling relation, while $\beta$ represents the normalization. The best-fitting power-law relation yields a scaling of $M \propto R^{1.41}$. The clump masses are estimated using Equation~\ref{eq:clump_mass}, which assumes virial equilibrium, yielding $M_{\rm vir}\propto R\sigma^{2}$.

Previous studies of massive star-forming clumps have reported a broad range of power-law slopes ($\alpha$) for the mass--size relation. Typical values are $\alpha \simeq 1.6$--$1.7$ \citep{Lombardi2010A&A...519L...7L,Kauffmann2010ApJ...716..433K,Urquhart2014MNRAS.443.1555U}, while steeper relations with $\alpha \gtrsim 2$ \citep{Ellsworth2015ApJ...805..157E} and even $\alpha \gtrsim 2.7$ \citep{Ragan2009ApJ...698..324R} have also been reported.

The slightly shallower slope obtained for our sample may reflect departures from virial equilibrium, variations in the internal density profiles of the clumps, or the influence of turbulence and stellar feedback. Nevertheless, the observed mass--size scaling is broadly consistent with that expected for self-gravitating systems, suggesting that the identified clumps are largely compatible with gravitationally bound star-forming regions.

\section{Discussion}
The XMP galaxies in our sample are largely isolated on scales of $\sim$ 0.5 Mpc, with no nearby potential perturbers \citep{Filho2015ApJ...802...82F}. Despite this isolation, their ionised gas kinematics reveal diverse behaviour. Only three galaxies show coherent velocity gradients consistent with ordered rotation, while the majority exhibit disturbed, asymmetric, or irregular rotation curves. This indicates that a significant fraction of the XMP galaxies in our sample are not dynamically relaxed.

In several galaxies, localized deviations from the ordered rotation coincide with compact, bright star-forming clumps. This behaviour is particularly evident in J0910+0711, SBS1415+437, and VCC0428, where the velocity field exhibits localized jumps at the clump boundaries superimposed on otherwise regular rotation curves. A counter-rotation with a maximum velocity of 6~km\,s$^{-1}$ at the position of the head has also been reported by \citet{SanchezAlmeida2013ApJ...767...74S}. These velocity perturbations are comparable to the clump velocity dispersions and likely reflect the additional gravitational influence of the clumps on the surrounding ionized gas. The coincidence of the velocity jumps with the clump edges suggests that these star-forming regions are dynamically significant, contributing to the local gravitational potential rather than simply tracing the global rotation of the host galaxy. The observed perturbations may also arise from a combination of effects, including local variations in the gravitational potential associated with massive star-forming clumps, turbulence driven by stellar feedback such as winds and supernovae, and unresolved multiple kinematic components linked to clustered star formation. This highlights that the ionised gas kinematics in XMP galaxies are shaped not only by global galaxy rotation, but also by significant small-scale dynamical substructure.

Beyond these localized effects, several systems exhibit more extended kinematic signatures, including declining rotation curves and counter-rotating or kinematically decoupled ionized-gas components. These features cannot be adequately described by a single rotating-disk model and suggest the presence of multiple kinematic components with distinct angular momentum orientations within the same system. Such kinematically distinct structures may arise from external gas acquisition, either through the accretion of relatively metal-poor gas from the circumgalactic or intergalactic medium or through interactions and mergers with gas-rich companions. If the gas was accreted near the star-forming head, it may not yet have fully mixed with the pre-existing disk, potentially producing the observed kinematic complexity. Interactions between distinct gaseous components can further generate velocity reversals, kinematic twists, and departures from simple axisymmetric rotation. These mechanisms have been widely proposed to explain kinematically decoupled gas in nearby galaxies \citep[e.g.,][]{Bertola1992ApJ...401L..79B, Thakar1996ApJ...461...55T, Thakar1997ApJ...479..702T, Corsini2014ASPC..486...51C, Yadav2021ApJ...914...54Y, Das2021JApA...42...85D, Yadav2026A&A...709A.172Y}.

Numerical simulations and observations indicate that smoothly accreted gas can produce rotational lags with respect to the pre-existing disk \citep{Fraternali2006MNRAS.366..449F, Ho2019ApJ...875...54H, Bizyaev2022MNRAS.515.1598B}. The formation of counter-rotating gas, therefore, requires the accreted material to retain substantially misaligned angular momentum, a condition that may be more readily achieved in low-mass galaxies owing to their relatively small pre-existing gas reservoirs. In such systems, cold-mode gas accretion is expected to be efficient, supplying fresh gas that can drive turbulence and promote the formation of massive star-forming clumps \citep{Keres2005MNRAS.363....2K, Dekel2006MNRAS.368....2D, Dekel2009Natur.457..451D, Fensch2021MNRAS.505.3579F}.

An alternative scenario involves the accretion of gas-rich dwarf companions. In this case, the acquired gas can preserve its original angular momentum and remain kinematically distinct over several dynamical times. Gaseous counter-rotation is widely interpreted as a signature of merger-driven acquisition of externally supplied gas with misaligned angular momentum \citep{Yadav2023MNRAS.526..198Y, Yadav2024A&A...689A.346Y, Beom2024AJ....168..197B}. The apparently isolated environments of the sample galaxies do not exclude this possibility, as low-mass companions may already have merged or been disrupted, leaving tidal features too faint to detect in such low-mass systems.

Similar evidence for externally driven gas perturbations has been reported in other isolated, low-density environments. Galaxies in the Void Galaxy Survey exhibit disturbed HI morphologies and kinematics consistent with ongoing interactions or gas accretion \citep{Kreckel2012AJ....144...16K}, while disturbed HI disks have also been identified in XMP void galaxies \citep{Kurapati2024MNRAS.533.1178K}. Localized non-circular ionised-gas motions associated with chemical abundance variations have likewise been reported in NGC~428 and Ark~18 \citep{Egorova2019MNRAS.482.3403E, Egorova2021MNRAS.504.6179E}. 

Galaxy interactions can also substantially modify the global kinematics by reducing the ratio of ordered to random motions (V/$\sigma$), with the strongest dynamical heating expected in major mergers \citep{Bournaud2005A&A...437...69B, Cox2006ApJ...650..791C}. The relatively low V$_{\textrm{H}\alpha}/\sigma_{V}$ values measured in the sample galaxies are therefore consistent with dynamically disturbed systems, although they do not uniquely distinguish between merger-driven evolution and external gas acquisition.

The role of stellar feedback may also contribute to the observed kinematics. Stellar winds and supernova explosions can generate turbulent motions, multiple velocity components, and broadened emission-line profiles \citep{McKee2007ARA&A..45..565M, Veilleux2005ARA&A..43..769V}. Such processes are likely responsible for part of the disturbed velocity fields seen in the sample galaxies. In particular, feedback-driven gas, including blueshifted and redshifted outflows or supernova ejecta, can naturally introduce locally distinct kinematic components that may add to or even contribute to counter-rotating signatures in some regions \citep{Nesvadba2008A&A...491..407N, Harrison2012MNRAS.426.1073H, Rupke2013ApJ...768...75R, Yadav2021A&A...651L...9Y}. However, these effects are typically associated with enhanced velocity dispersion and small-scale deviations from ordered rotation, whereas the counter-rotating galaxies in our sample show coherent velocity reversals extending over several hundred parsecs. Similarly, AGN-driven outflows can generate complex gas kinematics, including disturbed velocity fields, multiple velocity components, and enhanced velocity dispersions \citep{Izotov2008ApJ...687..133I, Fischer2018ApJ...856..102F,  Burke2021MNRAS.504..543B,
Yadav2026ApJ...998..267Y}. Together with stellar feedback from winds and supernovae, these processes may contribute to the observed kinematic complexity and influence the structure of the ionised gas.

The low metallicities measured in the star-forming clumps are broadly consistent with the low-metallicity regime occupied by low-mass dwarf galaxies and with the low-mass extension of the local mass--metallicity relation. In such low-mass systems, inefficient metal retention, stochastic star formation and enrichment, mixing, outflows, and recycling can all contribute to maintaining low gas-phase abundances. The long gas-depletion times typical of dwarf galaxies also mean that the observed star formation need not require continuous external gas supply.

External gas accretion nevertheless remains a plausible contributor to the evolution of these systems. In particular, the disturbed and kinematically distinct ionised-gas components observed in several galaxies may indicate a complex gas-supply history, although they do not uniquely establish the origin of the low metallicities. This interpretation is supported by HI studies of XMP galaxies, which reveal substantial neutral-gas reservoirs and, in some cases, disturbed HI morphologies and kinematics that have been associated with interactions or possible cold-gas accretion \citep{Ekta2008MNRAS.391..881E, Ekta2009MNRAS.397..963E, Ekta2010MNRAS.403..295E}. Consequently, the observed properties are likely to reflect an interplay between external gas supply and internal processes governing metal production, retention, recycling, and mixing \citep{SanchezAlmeida2014A&ARv..22...71S, Almeida2015ApJ...810L..15S, Cresci2010Natur.467..811C, Cao2022NatAs...6.1464C}. Spatially resolved observations of the neutral gas, together with deeper measurements of the ionised-gas metallicity and kinematics, will be required to constrain the relative contribution of these processes.

A key result is that the off-centre H$\alpha$-bright clumps are spatially associated with kinematic disturbances and velocity reversals, revealing a close connection between local gas dynamics and ongoing star formation. The presence of kinematically distinct ionized-gas components may indicate external gas acquisition, potentially associated with gas accretion or a merger. The interaction between distinct gaseous components can enhance gas compression, turbulence, and orbit crowding, thereby creating conditions favourable for star formation.

Taken together, the isolated environments, disturbed ionised-gas kinematics, dynamically significant star-forming clumps, and extremely low metallicities are difficult to reconcile with a simple closed-box evolutionary scenario. Instead, they are consistent with galaxies whose gas reservoirs are being replenished through external gas supply \citep{SanchezAlmeida2014A&ARv..22...71S, Almeida2015ApJ...810L..15S, Cresci2010Natur.467..811C, Cao2022NatAs...6.1464C}. The continued acquisition of fresh gas is expected to play a fundamental role in galaxy assembly by sustaining star formation and driving the gradual build-up of stellar mass \citep{Dekel2009Natur.457..451D, Silk2012RAA....12..917S, Ho2019ApJ...875...54H, Yadav2022A&A...657L..10Y}. In this context, the kinematically decoupled components, localized metal-poor star formation, disturbed velocity fields, and off-centred star-forming regions observed across our sample represent a coherent picture of ongoing growth in extremely metal-poor galaxies.

\section{Summary and conclusions}
\begin{enumerate}
    \item We analysed long-slit spectroscopic observations of nine XMP galaxies obtained with the INT/IDS instrument, focusing on the H$\alpha$ and [\ion{N}{ii}] emission lines.

    \item We derived ionised gas rotation curves by spatially binning the spectra, subtracting the continuum emission, and fitting Gaussian profiles to the H$\alpha$ emission line.

    \item The kinematic analysis reveals that only a subset of galaxies exhibit ordered rotation, while multiple systems show disturbed, asymmetric, or declining velocity profiles.

    \item Three galaxies show evidence for multiple kinematic components, including possible counter-rotating gas, suggesting complex ionised gas structures.

    \item We estimated the masses, effective radii, and metallicities of the dominant star-forming clumps and host galaxies. The derived clump mass--size relation is consistent with previous studies of XMP galaxies, with clump masses ranging from $10^{6}$ to $10^{9}$\, M$_{\odot}$.

    \item The measured metallicities confirm the XMP nature of the star-forming regions.

    \item The combination of disturbed kinematics, low metallicities, clumpy morphologies, and counter-rotating ionized gas disks supports scenarios in which external gas accretion drives star formation and influences the dynamical evolution of XMP galaxies.
\end{enumerate}

\begin{acknowledgements}
    We thank the referee for the valuable suggestion, which has helped us improve the manuscript. The authors acknowledge financial support from the Spanish Ministry of Science, Innovation and Universities (Ministerio de Ciencia, Innovación y Universidades, MICIU), project PID2022-136598NBC31 (ESTALLIDOS8) by MCIN/AEI/10.13039/501100011033. This work is part of grant CEX2025-001609-S, awarded to the Instituto de Astrofísica de Canarias under the Severo Ochoa Centre of Excellence program and funded by MICIU/AEI/10.13039/501100011033. This work is based on observations obtained with the Isaac Newton Telescope, operated on the island of La Palma by the Isaac Newton Group of Telescopes at the Spanish Observatorio del Roque de los Muchachos of the Instituto de Astrofísica de Canarias. The observations were carried out as part of the Spanish TAC Programme C90 during the nights of 16 and 17 February 2016.
\end{acknowledgements}

\bibliographystyle{aa} 
\bibliography{references}

%


\begin{appendix}\label{sec:appendix}
\onecolumn

\section{Approximate dynamical masses of the galaxies}\label{sec:dynamical_mass}
We estimated the dynamical masses from the rotation curves assuming an approximately spherical mass distribution. In this framework, the circular velocity (V$_{\mathrm{c}}$) at a given radius $d-d_{0}$ is determined by the mass enclosed within that radius, $M(d-d_{0})$, through the equilibrium between gravitational attraction and centrifugal support.

\begin{equation}
V_{\mathrm{c}}^{2}(d-d_{0}) = \frac{G\,M(d-d_{0})}{d-d_{0}}.
\end{equation}
where G is the gravitational constant.

The observed line-of-sight velocities correspond to the projected intrinsic circular velocities and are corrected for disk inclination using a factor of $\sin i$, i.e., $V_{\mathrm{H}\alpha} = V_{\mathrm{c}} \sin i$.

The enclosed dynamical mass within a given radius is inferred using the formula given by 

\begin{equation}
M(d-d_{0})\sin^{2} i = \frac{d - d_{0} \times \,V_{\mathrm{H}\alpha}^{2}(d-d_{0})}{G}.
\end{equation}

Using the value of the gravitational constant, the equation turns out to be

\begin{equation}
M(d-d_{0})\sin^{2} i = (2.33 \times 10^{5} M_\odot)\,(d - d_{0})\,V_{\mathrm{H}\alpha}(d-d_{0}),
\end{equation}

where $d-d_{0}$ is measured in kpc and $V_{\mathrm{H}\alpha}$ is in km\,s$^{-1}$. This approach provides a first-order estimate of the dynamical mass distribution. Additional effects, including pressure support, non-circular motions, and deviations from spherical symmetry, may contribute to the uncertainty in the dynamical mass estimates. In our sample, the observed $V_{\textrm{H}\alpha}/\sigma_{V}$ ratios indicate that asymmetric drift can provide a non-negligible contribution to the dynamical support. However, applying a robust asymmetric-drift correction requires well-constrained radial profiles of the gas surface density and velocity dispersion, which are not sufficiently sampled over the limited H$\alpha$ extent of our observations. Given the limited spatial extent of the observed H$\alpha$ emission, applying such a correction would introduce additional assumptions and uncertainties. We therefore do not apply an asymmetric-drift correction and regard the derived dynamical masses M$_{dyn \thinspace approx}$ based on observed H$\alpha$ extent R$_{H\alpha}$ as approximate estimates that may underestimate the true dynamical masses \citep{Read2016MNRAS.462.3628R, Oh2025Ap&SS.370..129O}.

\section{Two-component velocity curves}

\begin{equation}\label{eq:vel_component}
\begin{aligned}
    V(d\ |\ V_{0},V_1,d_0,\Delta) &= V_{0} + V_{1} \frac{d-d_{0}}{\sqrt{\Delta^{2}+(d-d_{0})^2}}
    \end{aligned}
\end{equation}

\begin{equation}\label{eq:flux_component}
\begin{aligned}
      f_1(d\ |\ p, d_0) 
      &= \exp{\left(-\frac{|d-d_{0}|}{p}\right)}
      ,\quad \text{AND} \quad p>0, \quad \text{For central region}\\
      f_2(d\ |\ p, d_0) 
      &= 1-f_1(d\ |\ p, d_0), \quad \text{For outer region}
\end{aligned}
\end{equation}

\begin{equation}\label{eq:combined_flux_weighted}
\begin{aligned}
    V_\text{unconstrained}^\text{flux weighted}(d) 
      &= \frac{f_1(d\ |\ p, d_{0,a}) V(d\ |\ V_{0,a},V_{1,a},d_{0,a},\Delta_{a}) 
      + f_2(d\ |\ p, d_{0,b}) V(d\ |\ V_{0,b},V_{1,b},d_{0,b},\Delta_{b})}
      {f_1(d\ |\ p, d_{0,a})+f_2(d\ |\ p, d_{0,b})}  
\end{aligned}
\end{equation}

\begin{equation}\label{eq:constrined_same_center_same_velocity}
\begin{aligned}
    V_\text{same centre, same velocity}^\text{flux weighted}(d) 
      &= V_\text{unconstrained}^\text{flux weighted}(d)
      ,\quad \text{AND} \quad V_{0,a}=V_{0,b}, d_{0,a}=d_{0,b} 
\end{aligned}
\end{equation}

\begin{equation}\label{eq:corotating_components}
\begin{aligned}   
    V_\text{co rotating}^\text{flux weighted}(d) 
      &= V_\text{same centre, same velocity}^\text{flux weighted}(d)
      ,\quad \text{AND} \quad V_{1,a}V_{1, b}>0 
\end{aligned}
\end{equation}

\begin{equation}\label{eq:counter_rotating_components}
\begin{aligned}  
    V_\text{counter rotating}^\text{flux weighted}(d) 
      &= V_\text{same centre, same velocity}^\text{flux weighted}(d)
      ,\quad \text{AND} \quad V_{1,a}V_{1, b}<0 
\end{aligned}
\end{equation}

The observed line-of-sight velocity at a projected distance $d$ along the slit is denoted by $V(d)$ (Equation~\ref{eq:vel_component}), the parameters $V_{0}$, $V_{1}$, $\Delta$, and $d_{0}$ are treated as free parameters and determined through the fitting procedure. Here, $V_{0}$ represents the systemic velocity, $V_{1}$ the velocity amplitude, $\Delta$ the velocity gradient of the component, and $d_{0}$ the kinematic centre of the rotation curve.

The corresponding flux distribution is described by Equation~\ref{eq:flux_component}, where $f_1(d)$ and $f_2(d)$ denote the flux at a projected distance $d$ along the slit with $p$, and $d$ as free parameters. One of the components is assumed to have an exponential profile while the total flux is assumed to be unity. Alternatively, the two gas-gas components' flux profiles could be modelled as exponential+doughnut, disk+doughnut, box function+inverse box function, and combinations of sigmoid functions. We chose the simplest model due to the paucity of available data and the reduction in free parameters.

The observed rotation curve is modelled using the flux-weighted combination of the two kinematic components given by Equation~\ref{eq:combined_flux_weighted}. In this formulation, the relative contribution of each component varies according to its flux. To investigate the nature of the observed kinematic substructure, we explored four model configurations. These include an unconstrained model, in which the two components are allowed to have independent kinematic centres and systemic velocities, and a constrained model (Equation~\ref{eq:constrined_same_center_same_velocity}), in which both components share a common kinematic centre and systemic velocity. In addition, both co-rotating (Equation~\ref{eq:corotating_components}) and counter-rotating (Equation~\ref{eq:counter_rotating_components}) configurations were considered. Among all the models examined, the constrained two-component model provides the best representation of the observed rotation curves and is therefore adopted in this work.

\end{appendix}
\end{document}